# Experimental and theoretical investigation of angular dependence of the Casimir force between sinusoidally corrugated surfaces

A.A. Banishev[1], J. Wagner[1], T. Emig[2], R. Zandi[1], and U. Mohideen[1]

[1]Department of Physics and Astronomy, University of California, Riverside, California 92521, USA

[2]Laboratoire de Physique Théorique et Modéles Statistiques, CNRS UMR 8626, Université Paris-Sud, 91405 Orsay, France

In the current work we present the complete results for the measurement of normal Casimir force between shallow and smooth sinusoidally corrugated gold coated sphere and a plate at various angles between the corrugations using an atomic force microscope. All measured data were compared with the theoretical approach using the proximity force approximation and theory based on derivative expansion. In both cases real material properties of the surfaces and non-zero temperature were taken into account. Special attention is paid to the description of electrostatic interactions between corrugated surfaces at different angles between corrugations and samples preparation and characterization. The measured forces are found to be in good agreement with the theory including correlation effects of geometry and material properties and deviate significantly from the predictions of the proximity force approximation approach. This provides the quantitative confirmation for the observation of diffraction-type effects that are disregarded within the PFA approach. The obtained results open new opportunities for control of the Casimir effect in micromechanical systems.

## I. INTRODUCTION

The Casimir effect [1] has become well known due to many potential applications in both fundamental physics and nanotechnology. The most familiar is the attractive Casimir force between two planar neutral conducting surfaces placed in vacuum. Because the Casimir force goes inversely as a large power of the distance between surfaces it is large at sub-micron separations and plays an important role in micro- and nano-electromechanical systems (MEMS and NEMS). In MEMS/NEMS devices a common failure mode is the collapse of neighboring surfaces onto each other or the jump to contact of moving components with adjacent surfaces due to the Casimir force. This phenomenon is usually called stiction, pull-in effect or snap-down effect and has been a serious problem in NEMS/MEMS operation [2-7]. In condensed matter physics the Casimir effect finds application in the study of the properties of thin films and critical phenomena [8,9]. The precision measurement of the Casimir force has also been advanced as a powerful test for proposed hypothetical long-range interactions, including corrections to the Newtonian gravitational law at small distances predicted by the unified gauge theories, supersymmetry, supergravity and string theory [10-12]. Hence, there is an important need for further research on the Casimir effect motivated by the fact that it is finding new applications in both fundamental science and engineering.

The Casimir effect is viewed as originating from the modification of the quantum vacuum photon fluctuation spectrum due to the presence of boundaries such as the parallel plates. This approach naturally suggests a strong dependence of the force on the shape of the boundary. Many intriguing shape dependences have been predicted, including the possibility of obtaining repulsive forces for ideal metal spheres [13-14] and cubes [15-17]. These exotic shape dependences are yet to be tested. Uniformly corrugated surfaces provide a more convenient platform to explore some key aspects of the shape dependent Casimir force such as coherent diffraction like scattering of zero point photons [18-27].

Alternately, the Casimir force can be viewed as the collective interactions of the charge and current fluctuations induced by the photons of the quantum vacuum on the two bodies. At non zero temperature, there is also a thermal photon contribution. Uniformly corrugated surfaces are an ideal system to explore the interplay of boundary shape and the length scale of the charge and current fluctuations. The coupled geometry and material dependence of the Casimir force can be further enhanced by a measurement using two corrugated surfaces. The forces between the two corrugated surfaces can be studied with their axis aligned but as a function of the phase shift [21-24] or as a function of angle between the corrugation axis [28]. For understanding the role of coherent diffraction like effects, the important experimental size scales are corrugation period $\lambda$, and the separation between the corrugations, $z$. The vacuum and thermal photon wavelengths of interest are those that correspond to separation distance and the thermal wavelength $\lambda_T=\hbar c/k_B T$. These couple with those representing the material reflectivity through the plasma wavelength $\lambda_p=2\pi c/\omega_p$ and the free electron scattering length $\lambda_\gamma= 2\pi c/\gamma$. The interplay of the different length scales associated with the photon wavelengths and boundary reflectivity, with the corrugation period and the angle between the two corrugations lead to a rich behavior in this system, making it a promising probe of these coupled phenomena [29-32]. The strong observed dependence of the Casimir force on the corrugation angle means, that this feature can be used in adjusting and controlling the moving parts in proposed micromechanical devices using corrugated surfaces and the Casimir effect [33-35].

The normal Casimir force acting in the direction perpendicular to the interacting surfaces has been the most investigated aspect of the Casimir effect. This force component was studied between smooth surfaces and measured using various techniques such as the spring balance [36,37], the torsion balance [38,39] the AFM [40-46], macroscopic oscillators [47] and the micro torsion oscillator [4,5,48,49]. These studies have highlighted the material dependence of the Casimir force. Agreement between the measured data is obtained only when the material properties are taken into account. A key question on the role of free electron dissipation remains to be understood [39,44,49-58]. The optical modulation of the normal Casimir force has been demonstrated [59] and optically transparent boundaries have been used to cancel the Casimir force [60].

The normal Casimir force has also been studied between a corrugated surface and a large spherical surface using the AFM [20] and a microtorsional oscillator [25-27]. These studies have pointed out the strong deviation of the measured Casimir force from approximate approaches such as the Proximity Force Approximation (PFA) [61,62]. In the simple PFA, opposing curved surfaces are then treated as infinitesimal parallel plates and Casimir energy is found as an additive sum of the corresponding local parallel plate energies. But Casimir forces are non-additive and the PFA neglects diffraction effects from the curved boundaries and correlations from the interplay of geometry, material properties and temperature. In addition, 1-D periodic structures such as corrugations modify the collective coherence effects. The effect of using two periodically corrugated surfaces has also been studied using the lateral Casimir force [19,21-24,63]. In the lateral Casimir force the force tangential to the two surfaces, induced by a phase shift between the two aligned periodic corrugations is studied. For sinusoidal corrugation periods on the order of the separation distance a strong deviation from PFA including asymmetric force profiles where observed.

The problem of a precise description of the Casimir effect and the surface geometry, for two interacting corrugations, remains nontrivial, which stimulates further investigation of geometry dependence of the Casimir effect. The use of two corrugations allows the additional parameter of the orientation angle between them as a means to study the strongly coupled geometry and material dependence. Changing the corrugation orientation angle modifies the effective length of the fluctuations. As remarked above sinusoidal corrugations when made of real materials is of special interest because provide an additional system to better understand the macroscopic effects of vacuum fluctuations and the coupling between the material dependent characteristic length scales. The experimental exploration of this problem might also be helpful for clarifying how to simultaneously consider both the dielectric properties of the interacting materials and their deformed geometry with sufficient precision, as well as, for estimation of uncertainties which can arise due to using the approximation methods in the Casimir force calculations. In this configuration both a normal and a lateral Casimir force can be detected. The rotation in the orientation of the two corrugations has also been proposed as a mechanism to generate vacuum fluctuation induced torques [64].

In this paper we present the full description of experiment and theory of the normal Casimir force acting between a shallow and smooth sinusoidally corrugated sphere and a plate covered with a gold layer at different crossing angles between corrugations. Some of the results were briefly described in Ref. [28]. We present the description of experimental procedure and the measurement in more complete detail. In addition details of the theory underlying the perturbative computation of electrostatic force and the derivative expansion of the Casimir force are included. The forces were measured for corrugation periods of 570.5 nm and crossing angles from 0 to $2.4^o$. The measured forces at 300 K are compared with the theory based on the derivative expansion including the material properties with no fitting parameters. The derivative expansion for smooth surfaces of the Casimir free energy is a local expansion in terms of the gradient of the height profile of the surfaces, regardless of the amplitude to distance ratio. It provides the leading order correction to the PFA. It is shown that inside the experimental error interval of 67% confidence level the measured Casimir forces are in agreement with the derivative expansion. For the corrugation wavelength used, the Casimir force increases by 15% at the closest sphere-plate separation point of 130 nm when the crossing angle between corrugations goes from 0 to $2.4^o$. The experimental data are also compared with theory based on PFA applied to both the corrugations and the curvature of the sphere. The material properties are included through the dielectric function as in Eq. (28). The PFA approximates the force by $F^{PFA} = 2\pi R\, U_{corr}^{PFA}$, where $U_{corr}^{PFA}$ is the PFA approximation to the Casimir free energy per unit area between two corrugated but otherwise planar surfaces as given by the Lifshitz theory, see Eqs. (15), (18) below. Strong deviation from the PFA theory is observed, pointing to the important role of geometry even for the small and smooth corrugation amplitudes used.

The paper is organized as follows. In Sec. II the experimental setup, preparation and characterization of the samples are described. In Sec. III we describe the theory and experimental procedure for electrostatic calibration. In section IV the experimental results of Casimir force measurements, including the measurement errors are presented. In section V the derivative expansion and PFA theory that describes the normal Casimir force for the configuration of gold-coated sphere and a plate covered with sinusoidal corrugations is presented. In section VI the theory developed is compared with the measurement results. We end with the conclusion in Section VII.

**II. EXPERIMENTAL SETUP AND SAMPLE CHARACTERIZATION**

The experimental apparatus used in this study is described elsewhere [46]. Here, we provide an abridged description and refer readers to Ref. [46] for additional details. An exception is made with regard to the force sensor, which is different from the one used previously [46]. A schematic diagram of the experimental setup is shown in Fig. 1.

A standard AFM was used for the force measurements. It was placed in oil-free high vacuum chamber at pressure below $10^{-6}$ Torr and room temperature. The calibration of AFM piezo transducer movement was done with a fiber interferometer described elsewhere [65]. The AFM was tuned to work in contact mode for measuring vertical cantilever deflection for every 0.2 nm movement of the piezo actuator. For reducing mechanical noise, the system was maintained on an optical table and a sand damper box to prevent coupling of the low-frequency noise from the mechanical and turbo pumps was used. Liquid $N_2$ cooling was used to further lower the noise and to stabilize the laser power in the AFM.

To perform the Casimir force measurement, two aligned corrugated surfaces were required. This was achieved by imprinting a gold coated sphere onto a grating template as discussed below. Diffraction gratings with uniaxial sinusoidal corrugations and a 300 nm Au coating (from Horiba Jobin Yvon) was chosen as the first surface. The diffraction gratings were made on Pyrex glass and had a 300 nm Au coating. The same diffraction grating was used as a template for imprinting the corrugations on the top of the sphere as it described below. The grid surface was examined by an AFM and found to have a very homogeneous sinusoidal corrugation with a period of $\Lambda=570.5\pm0.2$ nm and an amplitude $A_1=40.2\pm0.3$ nm. The 3-D surface topography of the grating as measured with the AFM is shown in Fig. 2a. A minor sub-oscillation of 1-2 nm amplitude with a stochastic mean and 110-140 nm period resulting from the manufacturing is also observed. In Fig. 2b a typical cross section of the topography in Fig. 2a is depicted. These measurements were made after acquisition of the force data.

A ~1×1 cm$^2$ size of the diffraction grating was gently cut using a circular diamond cutter and the surface was checked for attached debris using an optical microscope. Then additional cleaning of the grating sample was done using the following procedure. First, the grating sample was sonicated in sulfur-free soap water for 10 min. After this, the piece was rinsed with DI water to remove soap and again sonicated in methanol and ethanol for 10 minutes each respectively. During the sonication process care was taken to delicately hold the sample far away from the container boundaries. Finally, it was dried with pure nitrogen and again examined with an optical microscope to check for surface damages. After confirming that the grating sample surface does not contain microscopic damages, it was fixed on a rotation stage. The stage with the sample grating was mounted on the top of the AFM piezo scanner (see Fig. 1). To provide electrical contact to the grating sample a thin copper wire was soldered to the edge of the grating using indium wire as a solder. This grating sample was used as a template for *in situ* imprinting of the corrugations on the bottom surface of an Au coated sphere of a specially prepared cantilever.

The force sensor was prepared in the following way. A polystyrene sphere of radius 100 μm was attached to the tip of a 320 μm long triangular silicon-nitride cantilever with a nominal spring constant of order 0.01 N/m using conductive silver epoxy. Using the cantilever with triangle configuration allowed us to suppress the lateral Casimir effect that can lead to the torsional deflection (rotation) of the cantilever. The torsional spring constant of triangle cantilever is much larger than that corresponding to the bending making it sensitive to detecting the normal Casimir force, while simultaneously suppressing the effect of the lateral Casimir force. To improve the adhesion of the sphere, the cantilever with the freshly attached sphere was placed under heat lamp (250 Watt) at the distance of about 10 cm for 30 min and then placed in vacuum chamber for 24 hours to let all volatile gas molecules to evaporate. After the sphere was secured, the cantilever-sphere system was coated with a 10 nm Cr layer followed by a 20 nm Al layer and finally with a 110±1 nm Au layer using oil-free thermal evaporator at a $10^{-7}$ Torr vacuum. To provide an uniform coating of the metals, rotation of the sensor during the coating process. The radius of the Au-coated sphere was determined using a SEM to be $R$=99.6±0.5 μm. After the sensor preparation was complete, it was inserted into the AFM cantilever holder and the AFM was placed inside the vacuum chamber for force measurements.

Next the *in situ* printing of the corrugations on the sphere bottom using the fixed diffraction grating on the sphere was done. After the chamber pressure reached $10^{-7}$ Torr, the grating sample was moved using a stepper motor to just touch the bottom of the sphere. The whole process was visually monitored, using a telescope and CCD camera attached to its output. The image was displayed on a large screen to precisely monitor the moment of sphere-plate contact. A metal stylus with a rounded end was slowly approached to the top side of the sphere using second stepper motor to gently touch the sphere (see Fig. 1). Then the AFM piezoelectric tube was extended to its maximum length. As a result, the sphere was squeezed between the grating and the stylus end leading to the imprint of the corrugations onto the sphere bottom. It was confirmed that the radius or ellipticity of the sphere remained unchanged. After the imprinting process, the metal stylus was removed and the force measurements were started. The topography of the imprinted corrugations measured using an AFM after completion of the force measurements is shown in Fig. 3a. In Fig. 3b, the profile of the corrugations perpendicular to the axis is fit to a sinusoid and the amplitude was found to be $A_2$=14.6±0.3 nm. The amplitude was relatively uniform as shown. A scanning electron image of the imprinted corrugations is shown in Fig. 4. The size of imprinted area was measured to be about $Lx \sim Ly \sim 14$ μm.

Special attention was paid to the evaluation of the inhomogenuities (roughness) of the corrugations on the grating sample and imprinted sphere. Here we use the procedure described in Ref. [24]. The stochastic roughness was obtained from AFM topography measurements with the same procedure for both surfaces. Here, we compared the measured surface profile of the corrugations (circles in Fig. 2b and 3b) with a sine function (solid line in Fig. 2b and 3b). Following this, we calculated the difference between the experimentally measured data points and the sine function and the corresponding rms deviation between the two. This was repeated at 20 different corrugation periods and the variance describing the stochastic roughness was found to be equal to $\delta_1 = 2.9$ nm and $\delta_2 = 1.9$ nm for the corrugated plate and sphere respectively.

For changing the orientation angle between the corrugations, the corrugated plate was rotated using a stage controlled with a stepper motor. The stepper motor was actuated by rectangular pulses from a function generator and the control of the pulses was monitored using an oscilloscope. Prior to the measurements, the stepper motor was independently calibrated and the uncertainty in the rotation angle was determined to be 0.1°. The Casimir forces between the corrugations were measured at the crossing angles of $\theta = 0$, 1.2, 1.8 and 2.4 degree respectively.

### III. ELECTROSTATIC INTERACTION

The deflection of the cantilever in response to a force between the corrugated sphere and plate is calibrated using the electrostatic force. The calibration allows the determination of the values of such parameters as the cantilever spring constant $k$, average separation distances on contact $z_0$ and the residual potential difference $V_0$ between the sphere and the corrugated plate. These parameters are obtained as a result of comparison of experimentally measured electrostatic force with a theoretical model of electrostatic interaction between sphere and plate.

The total force between the corrugated sphere and plate is given by the sum of electric and Casimir forces. The cantilever deflection signal due to the total force can be represented in the form:

$$S_{def}(z) = \frac{F_{tot}(z)}{\sigma'} = \frac{F_{el}(z) + F_{Cas}(z)}{\sigma'} = \frac{X(z)}{\sigma'}(V_i - V_0)^2 + \frac{F_{Cas}(z)}{\sigma'} \quad , \qquad (1)$$

where $F_{Cas}$ and $F_{el}=X(z)(V_i-V_0)^2$ are the Casimir and electrostatic forces between the corrugated sphere and plate, which are the functions of sphere-plate separation $z$. Here $z$ is measured from the mean values of the two corrugations. Even when both corrugations are grounded there is a residual potential $V_0$ which is the present between the surfaces due to the different surface work functions of the sphere and plate materials. This value of $V_0$ can result from the different paths taken to the ground, the polycrystalline nature of the Au coating or contaminants. The expression for the coefficient $X(z)$ is discussed below. The term $\sigma' \equiv \sigma m$ is the calibration constant of the cantilever measured in the units of force per unit deflection (pN/mV), where $\sigma$ is the cantilever spring constant and $m$ is the cantilever deflection in units of nm per unit photodetector signal.

## A. Theory

The coefficient $X(z)$ of theoretical electric force in Eq. (1) can be obtained in the following way. We employ a perturbative expansion to compute the electrostatic energy per area $A$ between two corrugated plates located at $z = H_1(\mathbf{r})$ and $z = H_2(\mathbf{r})$. It is given by

$$U_{el} = \frac{\varepsilon_0}{2A} \int_A d^2\mathbf{r} \int_{H_1(\mathbf{r})}^{H_2(\mathbf{r})} dz (\nabla \Phi)^2, \qquad (2)$$

where the potential $\Phi$ obeys

$$\nabla^2 \Phi = 0, \quad \Phi\big|_{z=H_1(\mathbf{r})} = 0, \quad \Phi\big|_{z=H_2(\mathbf{r})} = V, \qquad (3)$$

such that the surfaces have a voltage difference of $V \equiv V_i - V_0$. Solving Laplace's equation perturbatively in the height profiles yields the general expression for the energy of two surfaces $H_1(\mathbf{r}) = h_1(\mathbf{r})$ and $H_2(\mathbf{r}) = z + h_2(\mathbf{r})$ to second order in the height profiles $h_j(\mathbf{r})$,

$$U_{el} = \frac{\varepsilon_0 V^2}{2} \frac{1}{z} + \frac{\varepsilon_0 V^2}{z^2} \frac{1}{A} \int \frac{d^2\mathbf{k}}{(2\pi)^2} \left[ \frac{k}{2} \coth(kz) \left( |\tilde{h}_1|^2 + |\tilde{h}_2|^2 \right) - \frac{ke^{-kz}}{1-e^{-2kz}} \left( \tilde{h}_1 \tilde{h}_2^* + \tilde{h}_2 \tilde{h}_1^* \right) \right], \qquad (4)$$

where $\mathbf{k}$ is the in-plane wave vector of the Fourier transformed height profiles $\tilde{h}_j(\mathbf{k})$. We have assumed that the spatial average of $h_j(\mathbf{r})$ vanishes.

To study two corrugated surfaces, which have amplitudes $A_1$ and $A_2$, corrugation wavelength $\Lambda$ and crossing angle $\theta$ between the corrugation axes, we consider the two profiles

$$h_1(\mathbf{r}) = A_1 \cos(2\pi x / \Lambda), \\ h_2(\mathbf{r}) = A_2 \cos(2\pi(x \cos\theta - y \sin\theta)/\Lambda), \tag{5}$$

so that the Fourier transforms of the profiles are proportional to $\delta(\mathbf{k}-\mathbf{k}_j)+\delta(\mathbf{k}+\mathbf{k}_j)$ with $\mathbf{k}_1 = (2\pi/\Lambda)\hat{x}$ and $\mathbf{k}_2 = (2\pi/\Lambda)\hat{x}\cos\theta - (2\pi/\Lambda)\hat{y}\sin\theta$. Hence one has the integrals

$$\int \frac{d^2k}{(2\pi)^2}\left(|\tilde{h}_1|^2 + |\tilde{h}_2|^2\right) = 2\pi^2(\delta(\mathbf{k}_1-\mathbf{k}_1)+\delta(\mathbf{k}_2-\mathbf{k}_2)), \\ \int \frac{d^2k}{(2\pi)^2}\left(\tilde{h}_1\tilde{h}_2^* + \tilde{h}_2\tilde{h}_1^*\right) = 2\pi^2\delta(\mathbf{k}_1-\mathbf{k}_2), \tag{6}$$

where the delta functions for a plate of finite area $A = L_x L_y$ are given by

$$2\pi^2 \delta(\mathbf{k}_1-\mathbf{k}_1) = L_x L_y, \tag{7}$$

and

$$2\pi^2 \delta(\mathbf{k}_1-\mathbf{k}_2) = \int_{-L_x/2}^{L_x/2} dx \int_{-L_y/2}^{L_y/2} dy\, e^{i\frac{2\pi}{\Lambda}[x(1-\cos\theta)-y\sin\theta]} = \frac{\sin(\pi L_x(1-\cos\theta)/\Lambda)}{\pi(1-\cos\theta)/\Lambda} \frac{\sin((\pi L_y \sin\theta)/\Lambda)}{(\pi\sin\theta)/\Lambda}, \tag{8}$$

Using Eqs. (4), (6), (7) and (8), we get for the electrostatic energy per area for small values of $\theta$

$$U_{el} = \frac{\varepsilon_0 V^2}{2}\frac{1}{z} + \frac{1}{2}\frac{\varepsilon_0 V^2}{z^2}\left[\frac{\pi}{\Lambda}(A_1^2+A_2^2)\coth(2\pi z/\Lambda) - \frac{4\pi A_1 A_2}{\Lambda}\frac{e^{-2\pi z/\Lambda}}{1-e^{-4\pi z/\Lambda}}\frac{\sin(\pi L_y \theta/\Lambda)}{\pi L_y \theta/\Lambda}\right], \tag{9}$$

Eq. (9) was compared with a numerical computation of the electrostatic force using a finite element method for separations between 160 to 400 nm and for angles between 0° and 2.4°, and was shown to agree to better than 1%.

The usual PFA yields for the electrostatic force between a flat plate and a sphere of radius $R$ the result

$$F_{el}^{PFA} = 2\pi R U_{el}, \tag{10}$$

assuming that $R \gg z$. Since in the present experiment $R$ is also much larger than all other geometric length scales, we can apply Eq. (10) to the corrugated surfaces and combine it with Eq. (9) to obtain the coefficient $X(z)$ in Eq. (1). For small values of $\theta$ it becomes

$$X(z) = \frac{\pi\varepsilon_0 R}{z} + \frac{\pi\varepsilon_0 R}{z^2}\left[\frac{\pi}{\Lambda}(A_1^2 + A_2^2)\coth(2\pi z/\Lambda) - \frac{4\pi A_1 A_2}{\Lambda}\frac{e^{-2\pi z/\Lambda}}{1-e^{-4\pi z/\Lambda}}\frac{\sin(\pi L_y \theta/\Lambda)}{\pi L_y \theta/\Lambda}\right]. \tag{11}$$

where PFA requires $z \ll R$ and perturbation theory assumes $A_1, A_2 \ll z, \Lambda$. The ratio $z/\Lambda$ can be arbitrary.

### B. Experimental calibration

The calibration parameters were independently obtained for each value of crossing angle $\theta$ using the electrostatic force. For that purpose, the following measurement procedures were done. For application of voltages to the corrugated plate it was connected to a voltage supply operating with 1 µV resolution. To protect the sample surfaces from current surges when the surfaces come in contact, a 1 kΩ resistor in connected in series with the voltage supply. The cantilever with the attached corrugated sphere is grounded. To eliminate the adverse effect of electrical ground loops all the ground connections were unified and tied to the AFM ground.

The electrostatic force between corrugated plate and sphere as a function of the separation $z$ is measured for eleven different voltages $V_i$ applied to the corrugated plate. A range of voltages from -40 to -145 mV were applied to the corrugated plate. The mean separation $z$ between the bottom of the sphere and the corrugated plate is varied by applying voltages to the AFM piezo. For this purpose a 0.05 Hz continuous triangular voltage was applied leading to the piezo extension represented by $z_{piezo}$. Prior to all measurements, $z_{piezo}$ was calibrated interferometrically [65]. Starting at a maximum separation of 2 µm, the corrugated plate was moved towards the sphere and the cantilever deflection recorded every 0.2 nm. The net change in the corrugated sphere-plate separation is a sum of that from the piezo and the small contribution from the cantilever deflection and is given by [46]

$$z = z_{piezo} + mS_{def} + z_0, \qquad (12)$$

where, $mS_{def}$ is the change in separation distance due to cantilever deflection and $z_0$ is the average separation on contact of the two corrugated surfaces. All distances are referenced to the mean value of the corrugations. Although, cantilever deflections were acquired every 0.2 nm of $z_{piezo}$, the data analysis was done only for interpolated values at every 1 nm step.

After the deflection $S_{def}$ due to total forces were measured, the first step was to subtract any mechanical drift of the photodetector system with respect to the cantilever. For distances larger than 1.7 µm, the force between the Au sphere and grid is below the instrumental sensitivity. At these separations, the noise is larger than the signal and in the absence of systematic errors the signal should average to zero and have no dependence on the corrugated sphere-plate separation. Therefore, any linear alteration in signal $S_{def}$ is due to the mechanical drift of the cantilever-photodiode system. Such a linear drift was present even in the absence of the corrugated sphere and plate. To subtract this systematic drift the following procedure outlined in Ref. [66] was used. $S_{def}$ at distances larger than 1.7 µm was fit to a straight line, and the straight line was subtracted from the measured $S_{def}$ at all separation distances to correct for the effects of photodiode mechanical drift. This subtraction led to the mean deflection signal at large distances being equal to zero. This procedure was repeated for all experimental measurements. The next step was to precisely determine the point of corrugated sphere–plate contact and the cantilever deflection coefficient $m$ as described in Ref. [66]. The value of $m$ was determined to be 102.1±0.5 nm/unit deflection signal. The obtained value of $m$ was used to calculate the change in separation $mS_{def}$ due to the cantilever deflection. This with Eq. (12) determines the corrugated sphere-plate separation $z$ up to the value of $z_0$ (which is constant for the complete set of measurements).

The same electrostatic force dataset were used for determining the residual contact potential $V_0$, the cantilever calibration constant $\sigma'$, and the average separation on contact $z_0$. The parabolic dependence of the electrostatic force (Eq. (1)) on the applied voltage $V_i$ for a fixed separation $z$ was used in the determination of these quantities. The first step in the process is the determination of $V_0$ and the parabola curvature $\beta(z) \equiv X(z)/\sigma'$ at every corrugated sphere-plate separation $z$. An example is shown in Fig. 5 where the deflection signal $S_{def}$ measured at the corrugated sphere-plate separation $z_a = 135$ nm as a function of the applied voltage is shown as squares. This dependence was best $\chi^2$-fitted with parabolas (Fig. 5, line) to determine the value of $V_0$, the parabola vertex, and the value of the coefficient $\beta(z) \equiv X(z)/\sigma'$, the parabola curvature. The curvature of the parabola depends on the cantilever calibration constant $\sigma'$ and the average separation on contact $z_0$. The same procedure was repeated at each separation and $\beta(z)$ obtained as a function of $z$. Some values of $\beta(z)$ for $\theta = 0°$ are 14.86±0.06, 13.86±0.06, 8.06±0.06, 4.26±0.06 $V^{-1}$, at separation distances of 145, 155, 265 and 500 nm respectively. From best $\chi^2$ fitting of the experimentally obtained $\beta(z)$ by $X(z)/\sigma'$, where $X(z)$ is determined by Eq. (11), we obtained the values of $\sigma'$ and $z_0$.

It is important to check if $V_0$ changes with the corrugated sphere-plate separation $z$. Fig. 6a-d shows the $V_0$ obtained at all separations $z$ for the four different crossing angles used in the experiment. The presence of contaminants on the corrugated sphere and plate would lead to $V_0$ changing systematically with the separation $z$ [67-72]. The $V_0$ is found to be independent of separation and crossing angle. To check for possible systematic errors in the determination of $\sigma'$ and $z_0$, the following procedure was done [66]. First, the experimental $\beta(z)$ was fitted from the closest corrugated sphere-plate position to an end point $z_{end} = 1000$ nm, and the values of $z_0$ and $\sigma'$ were determined. Then the end point was decreased by 100 nm and the fitting was repeated, i.e. smaller range of $z$ values were used in the fitting procedure. That was repeated for 13 values of the end point each less than the previous by 100 nm at large $z_{end}$ and 50 nm at small $z_{end}$. The dependences of $\sigma'(z_{end})$ and $z_0(z_{end})$ are shown in Fig. 7-8 for all crossing angels $\theta$. Both are constant within the random errors and independent of the value of $z_{end}$ chosen. The independence of these two parameters on separation indicates the absence of separation distance calibration and other uncontrolled systematic errors. The same distance independency of calibration parameters were observed for all crossing angles $\theta$. From the value of $z_0$ the absolute separation distance can be determined and the value of $k'$ is used to convert $S_{def}$ to a force for each $\theta$ separately. The corresponding mean values of the parameters that were used for force calibration with the errors at 67% confidence level are given in the Table I.

## IV. CASIMIR FORCE MEASUREMENT

Using the cantilever calibration parameters above, the Casimir force was *calculated* from the subtraction of the electrostatic force from the total measured force Eq. (1) as $F_{cas}(z) = \sigma' S_{def}(z) - X(z)(V_i - V_0)^2$, where $V_i$ are the applied voltages to corrugated plate while the sphere remains grounded. We applied 11 voltages $V_i$ and at each voltage the cantilever deflection $S_{def}(z)$ was measured 10 times as a function of the corrugated sphere plate separation $z$.

This cantilever deflection $S_{def}(z)$ corresponds to that of the total force, Casimir and electrostatic. The mean value of $F_{Cas}$ as a function of $z$ (with a step of 1 nm) calculated from 110 individual values of the total force are shown as crosses in Fig. 9. The size of the cross corresponds to the horizontal and vertical total (random plus systematic) experimental errors at 67% confidence level. From Fig. 9 it can be observed that the magnitude of the Casimir force increases with the orientation angle. The Casimir force at a distance of 130 nm increases in the order 84.9, 88.8, 92.5 and 97.8 pN for orientation angles of 0°, 1.2°, 1.8° and 2.4° respectively for a total change of 15%. At a separation of 150 nm, the same forces are equal to 55.7, 57.8, 59.2 and 62.1 pN corresponding to a net increase of 11%. Note that this angle dependence is a finite size effect. For larger angles beyond 2.4°, the multiple crossings of the corrugations will lead to negligible angle dependence.

The error analysis of the experimental data was done as described in Ref. [44,46,73] for 67% confidence level. The variance of the mean value of Casimir force obtained from 110 measured force curves was found to be independent of corrugated sphere-plate separation. The mean values of the variance was found to be equal to 0.51, 0.45, 0.49 and 0.49 pN for the crossing angles 0°, 1.2°, 1.8° and 2.4° respectively. These values can be considered as random errors in the Casimir force measurements if we choose the 67% confidence level. For the 110 measurements, the degree of freedom is equal to 109. The systematic error in the measured forces is determined by [73] the instrumental noise (including the background noise) and errors in calibration. The latter is largely influenced by the errors in the measurement of the separation distances. Thus the systematic error is naturally separation dependent and increases at short distances. At the shortest separation, the maximum value of the systematic error in the Casimir force determination was found as 0.79 pN for $\theta=0°$ and the minimum as 0.73 pN for $\theta=1.8°$. To get the total absolute experimental error for the uncertainty in the measurement of the Casimir force, we quadratically added the random and systematic errors. An illustration of the typical dependence of the total error in the Casimir force determination on the corrugated sphere-plate separation is given in the Fig. 10 for the crossing angle of $\theta=1.2°$. The total error changes within the range from 0.88 to 0.78 pN as a function of the separation. In the same graph we plotted the random and systematic error dependences on the corrugated sphere-plate separation. The relative total experimental error in the Casimir force measurement increases with increasing the separation distance. For example for the separation distances $z=127$, 200 and 300 nm it was found to increase as (a) 1%, 3.2% and 10.3% for the $\theta=0°$, (b) 0.9%, 3.2% and 10.6% for the $\theta=1.2°$, (c) 0.9%, 3.2% and 10.1% for the $\theta=1.8°$, (d) 0.8%, 3.1% and 10% for the $\theta=2.4°$.

## V. THEORY OF CASIMIR FORCE BETWEEN CORRUGATIONS

To compare the experiment with the theoretical Casimir calculations, we need to take into account the geometric features, the finite temperature and real material properties. The important geometric features to consider are the sphere-plate geometry, the corrugations on both the sphere and the plate, and the angle of orientation between the two corrugated surfaces. In this approach, to apply the derivative expansion, the PFA is used to treat the curvature of the sphere, and relates the force to the energy per unit area as $F^{Der} = 2\pi R U_{corr}$, where the energy of two corrugated plates, $U_{corr}$, is calculated using the derivative expansion introduced by Fosco *et al.* [74] for scalar fields and Bimonte *et al.* [61,75] for the electromagnetic field in the presence of perfect conductors and dielectric materials. The latter calculation takes into account the material properties and finite temperatures as well as the corrugations. We expect that a first order derivative expansion is sufficient since the derivative of the surface profiles is of order 14.6/570.5 ~0.026 and 40.2/570.5 ~0.070, respectively.

Consider two almost flat periodic surfaces separated by an average distance $z$. Let $h_1(x,y)$ and $h_2(x,y)$ be the position dependent height profiles of the surfaces. The total local separation between the plates in the $z$ direction is then

$$H(x,y) = z + h_1(x,y) - h_2(x,y), \qquad (13)$$

Following [61], we can write the average energy per unit area between two slowly curving surfaces as

$$U_{corr} = \frac{1}{A}\int_A d^2x \left[ U(H) + \alpha(H)\nabla H \cdot \nabla H - \frac{1}{2}(HU'(H) - U(H))\nabla h_1 \cdot \nabla h_2 \right], \qquad (14)$$

where $A$ is the area of integration, $U(H)$ and $U'(H)$ are the Casimir energy per unit area between two *perfectly flat* parallel plates separated by a distance $H$ and its first derivative, and $\alpha(H)$ is a coefficient that is given by the derivative expansion (see below).

For two corrugated surfaces with an angle $\theta$ between the corrugations, the height profiles are explicitly given by Eq. (5). The first term in Eq. (14) corresponds to the traditional PFA. While there is some $\theta$ dependence for infinite sized systems due to the derivative terms in Eq. (14), a separate stronger $\theta$ dependence of the Casimir energy between corrugated plates can be found for finite sized plates. For any infinite sizes plates and any non-zero value of the angle $\theta$, the PFA energy per unit area between two plates is given by the integral over a unit cell,

$$U_{corr}^{PFA} = \frac{\sin\theta}{\Lambda^2} \int_0^\Lambda dx \int_{x\cot\theta}^{\frac{\Lambda + x\cos\theta}{\sin\theta}} dy\, U[(H(x,y))], \qquad (15)$$

where the prefactor is the inverse of the area of the parallelogram unit cell, and the limits of integration cover a single unit cell. Under the change of variables $x'=x$, $y'=x\cos\theta - y\sin\theta$, the distance $H(x,y)$ and the limits of integration become independent of $\theta$, and the Jacobian of the transformation cancels the $\sin\theta$ from the area of the parallelogram. Therefore in order to obtain a $\theta$ dependence to leading order, we must consider that at least one of the plates is of finite extent.

Assuming that one of the corrugated plates has dimensions $L_xL_y$, by substituting the profile functions into Eq. (14), we see that the integral will be highly oscillatory in both $x$ and $y$ unless $L_y \sin\theta/\Lambda$ is of order unity. In the case where $L_y \sin\theta/\Lambda$ is much larger than 1, the integral given in Eq. (14) can be approximated by the integral over a single unit cell, and for the same reasons given above there will be no $\theta$ dependence. Therefore, we can assume that $L_y \sin\theta$ is of the same order as $\Lambda$. Since the plate is large enough to contain many periods ($L_y \gg \Lambda$), the angle must be very small ($\sin\theta \ll 1$). We can simplify the profile function $h_2$ by using the small angle approximation

$$h_2(x, y) = A_2 \cos(2\pi(x - y\theta)/\Lambda), \tag{16}$$

The final result of the Casimir energy per unit area between two corrugated surfaces is obtained using Eq. (14) where the limits of integration are explicitly given by

$$\frac{1}{A}\int_A d^2x \rightarrow \frac{1}{\Lambda L_y}\int_0^\Lambda dx \int_{-L_y/2}^{L_y/2} dy, \tag{17}$$

and the profiles functions are given by Eqs. (13), (5), and (16). Errors in $L_x$ have negligible effect, as the average over one period is the same as over many periods. We have computed the error in the force introduced by a 0.5 μm error in $L_y$ and found it to be 0%, 1%, 1.6% and 1.5% for angles of 0°, 1.2°, 1.8° and 2.4° respectively at a mean separation of 100 nm. The effects of finite temperature and material properties are contained in the functions $U(H)$, $U'(H)$, and $\alpha(H)$. The function $U(H)$ is given by the Lifshitz formula

$$U(H) = k_B T \sum_{n\geq 0}' \sum_p \int \frac{d^2k}{(2\pi)^2} \ln\left(1 - r_p(i\zeta_n, k)e^{-2H\kappa}\right), \tag{18}$$

where $k_B$ is Boltzmann's constant, $T$ is the temperature, $p$ is the polarization (in this paper for *TM* polarizations $p=1$ and for *TE* polarizations $p=2$), $\zeta_n$ is the $n^{th}$ Matsubara frequency, $r_p(i\zeta_n,k)$ are the Fresnel reflection coefficients, $k$ is the magnitude of the $\boldsymbol{k}$ vector, and $\kappa = \sqrt{\zeta^2/c^2 + k^2}$ is the Wick rotated wavenumber in vacuum. The function $U'(H)$ can be simply calculated as the partial derivative of Eq. (18) with respect to $H$. The coefficient $\alpha(H)$ can be calculated from the small $k$ expansion of the kernel $\tilde{G}(k)$ of the 2<sup>nd</sup> order perturbation theory in the height profile. It is given by [61]

$$\alpha(H) = \frac{1}{2}\frac{\partial^2}{\partial k^2}\tilde{G}(k)\bigg|_{k=0} = \frac{k_B T}{2}\sum_{n\geq 0}{}' \int \frac{d^2 k'}{(2\pi)^2}\frac{\partial^2}{\partial k_x^2}\left(f^{s1}(\zeta_n,\mathbf{k}',\mathbf{k}'+\mathbf{k}) + f^{s2}(\zeta_n,\mathbf{k}',\mathbf{k}'+\mathbf{k})\right)\bigg|_{k=0}, \qquad (19)$$

with

$$f^{s1}(\zeta,\mathbf{k},\mathbf{k}') = \sum_{p,p'} 2\frac{\kappa r_p(i\zeta,k)e^{-2H\kappa}}{1+r_p^2(i\zeta,k)e^{-2H\kappa}}\frac{\kappa' r_p(i\zeta,k')e^{-2H\kappa'}}{1+r_p^2(i\zeta,k')e^{-2H\kappa'}} B_{pp'}(\zeta,\mathbf{k},\mathbf{k}') B_{p'p}(\zeta,\mathbf{k}',\mathbf{k}), \qquad (20)$$

$$f^{s2}(\zeta,\mathbf{k},\mathbf{k}') = -\sum_p \frac{\kappa r_p(i\zeta,k)e^{-2H\kappa}}{1+r_p^2(i\zeta,k)e^{-2H\kappa}} B_{2,pp}(\zeta,\mathbf{k},\mathbf{k};\mathbf{k}'), \qquad (21)$$

The functions $B_{pp'}$ and $B_{2,pp}$ are obtained by expanding the scattering matrix perturbatively in the height field [61,77]. The material properties enter the calculation via the dielectric function $\varepsilon(i\zeta)$ that is included in both the Fresnel reflection coefficients $r_p$ and the functions $B_{pp'}$ and $B_{2,pp}$. For completeness sake they are given here explicitly. The reflection coefficients are

$$r_1(i\zeta,k) = \frac{\varepsilon(i\zeta)\kappa - \bar{\kappa}}{\varepsilon(i\zeta)\kappa + \bar{\kappa}}, \qquad (22)$$

$$r_2(i\zeta,k) = \frac{\kappa - \bar{\kappa}}{\kappa + \bar{\kappa}}, \qquad (23)$$

where $\bar{\kappa} = \sqrt{\varepsilon(i\zeta)\zeta^2/c^2 + k^2}$ is the Wick rotated wavenumber in the material. The first function $B_{pp}$ is given by

$$B_{pp'}(\zeta,\mathbf{k},\mathbf{k}') = d_1(\zeta,k) \cdot \begin{pmatrix} \dfrac{\mathbf{k}\cdot\mathbf{k}'}{kk'} + \varepsilon(i\zeta)\dfrac{kk'}{\bar{\kappa}\bar{\kappa}'} & \dfrac{\hat{z}\cdot\mathbf{k}\times\mathbf{k}'}{kk'} \\ \dfrac{\hat{z}\cdot\mathbf{k}\times\mathbf{k}'}{kk'} & -\dfrac{\mathbf{k}\cdot\mathbf{k}'}{kk'} \end{pmatrix} \cdot d_1(\zeta,k'), \qquad (24)$$

where the indices $p=1,2$ and $p'=1,2$ number the element of the matrix B, and

$$d_1(\zeta,k) = \sqrt{\varepsilon(i\zeta)-1}\begin{pmatrix} \dfrac{\bar{\kappa}}{\varepsilon(i\zeta)\kappa+\bar{\kappa}} & 0 \\ 0 & \dfrac{\zeta/c}{\kappa+\bar{\kappa}} \end{pmatrix}, \qquad (25)$$

The second functions $B_{2,pp}$ are for *TM*- and *TE*-polarizations respectively

$$B_{2,11}(\zeta,\mathbf{k},\mathbf{k};\mathbf{k}') = 2\frac{\varepsilon(i\zeta)-1}{(\varepsilon(i\zeta)\kappa+\bar{\kappa})^2}\left[\frac{\varepsilon(i\zeta)-1}{\varepsilon(i\zeta)\kappa'+\bar{\kappa}'}\left(\varepsilon(i\zeta)k^2k'^2 - \bar{\kappa}^2\frac{(\mathbf{k}\cdot\mathbf{k}')^2}{k^2}\right)\right.$$
$$\left. + 2\varepsilon(i\zeta)\frac{\kappa'+\bar{\kappa}'}{\varepsilon(i\zeta)\kappa'+\bar{\kappa}}\kappa(\mathbf{k}\cdot\mathbf{k}') + \varepsilon(i\zeta)\bar{\kappa}\zeta^2/c^2 + \bar{\kappa}(\kappa'-\bar{\kappa}')\right], \quad (26)$$

$$B_{2,22}(\zeta,\mathbf{k},\mathbf{k};\mathbf{k}') = 2\frac{(\varepsilon(i\zeta)-1)\zeta^2/c^2}{(\kappa+\bar{\kappa})^2}\left[\frac{\varepsilon(i\zeta)-1}{\varepsilon(i\zeta)\kappa'+\bar{\kappa}'}\left(\frac{(\mathbf{k}\cdot\mathbf{k}')^2}{k^2}-k'^2\right) + \bar{\kappa}-\bar{\kappa}'+\kappa'\right], \quad (27)$$

The theoretical computation of the Casimir forces is performed with realistic properties of Au at 300K. The dielectric function of Au was expressed using a 6-oscillator model for the core electrons and the Drude model for the free electrons, which on the imaginary frequency axis is given by

$$\varepsilon(i\zeta) = 1 + \frac{\omega_p^2}{\zeta(\zeta+\gamma)} + \sum_{i=1}^{6}\frac{f_i}{\omega_p^2+\zeta^2+\zeta g_i}. \quad (28)$$

For Au we use the plasma frequency $\omega_p=9eV/\hbar$, the relaxation frequency $\gamma=0.035eV/\hbar$, and the oscillator constants $g_i$ from Ref. [49]. Small roughness corrections were taken into account as described in Ref. [24].

## VI. COMPARISON BETWEEN EXPERIMENT AND THEORY

### Angular dependence

The comparison of the experimental data with theory is shown in Fig. 9. No fitting parameters are used in the comparison of theory and experiment. At the start of the experiment i.e. $\theta=0^o$, the corrugations on the sphere and plate are considered to be in perfect registry, with the valleys in the former directly above the peaks of the latter due to the *in situ* imprint procedure used [21,22,24]. This means that for $\theta>0^o$, the peaks on the two corrugations approach each other leading to an increase in the magnitude of the attractive force observed. Good agreement between experiment and theory is found for all the crossing angles between the corrugations. We illustrate the coupled geometry and material dependence of the Casimir force from the corrugated plate-sphere system in two different ways. In Fig. 11 the deviation from PFA is explored by plotting the ratio of the experimental data to the force obtained from PFA (corresponding to $U_{corr}^{PFA}$). The deviations at the shortest common separation 127 nm, where the relative experimental error are small, are 7.7%, 4.7%, 2.3%, 1.8% for the angels $\theta$=0, 1.2, 1.8 and 2.4° respectively. This is consistent with our theoretical computations which indicate that the magnitude of the deviation saturates at ~2% for crossing angles >2$^o$ at the smallest separation for these corrugation parameters. Note that these deviations are observed even with the shallow smooth small amplitude corrugations on the sphere used. Due to the size of the error bars at the larger separations, no definitive observations on the change with $z$ for the different crossing angles can be made. Alternatively, one can observe the role of the diffraction like correlation effects and the interplay of the material properties on the geometry of the periodic corrugations by comparing the difference force obtained by subtracting the theoretical PFA force (corresponding to $U_{corr}^{PFA}$ at 300 K) from the measured values. This is displayed in Fig. 12. Here the difference between the experimental data and the PFA is compared to the difference force between the derivative expansion and the PFA, corresponding to $U_{corr}$ - $U_{corr}^{PFA}$ both at 300 K. The error bars (at 67% confidence level) represent the data and the theoretical difference is represented by the solid line. For clarity of observation only data at every 3 nm separation are shown in the figure. One can observe that there is a significant deviation of the experimental data from the theory based on simple PFA which ignores correlation effects. The difference is a measure of the diffraction like correlation effects. For the separation of 130 nm the absolute deviation is 5.9, 4.2, 2.1, 0.98 pN for crossing angles $\theta$=0, 1.2, 1.8 and 2.4° respectively. Note that while the magnitude of the force (Fig. 9) increases with crossing angle the difference force has the opposite relationship. The solid lines in Fig. 12 which represents the deviation of the derivative expansion from PFA are in good agreement with the deviation from PFA observed in the experiment. The agreements show that the derivative expansion is a good approach for understanding the complete Casimir force between two corrugated surfaces.

The role of the diffraction like correlation effects can be understood in relative isolation from the material dependence by comparing the same difference of the measured force from PFA to the theoretical difference force computed by subtracting the PFA force from that obtained using the derivative expansion for ideal metal corrugated surfaces at 300K. The theoretical difference is shown as dotted line in Fig. 12. The difference between the dashed and solid lines is a measure of the theoretical material dependence. In Fig. 12(a) the ideal metal result deviates from the observed difference force. The apparent agreement in Fig. 12(b) is a numerical coincidence and is part of a trend, where the theoretical force difference for an ideal metal increases with crossing angle. The difference is more clear for the other angles as shown in Fig. 12(c)-(d). It should be noted that this is only a difference between forces, and the total Casimir force is always larger for the perfect metal. To explore the role of temperature the ratio of the experimental data to the force from the derivative expansion at 300 K and 0 K is shown in the inset to Fig. 12. The data is found to be consistent with 300 K particularly at the smaller crossing angles.

**VII. CONCLUSIONS AND DISCUSSION**

In conclusion, we have experimentally demonstrated the angle dependence of the normal Casmir force between a corrugated plate and corrugated sphere. An Au coated plate with sinusoidal grating of period 570 nm and an amplitude of 40.2 nm was used as one surface. A pressure imprinting procedure was used to transfer the corrugations to the bottom of Au coated sphere resulting in aligned corrugations with the peaks of the corrugated plate corresponding to valleys in the imprint on the sphere. The normal Casimir force was measured at different crossing angles between the corrugations which was achieved by rotating the corrugated plate. The residual potential and the mean separation on contact between the two corrugations were verified to be independent of separation. Both random and systematic errors were found and combined to give a total error, which was used in the comparison. The random error was found to be independent of separation between the corrugations. The systematic error increased at smaller separation due to the distance dependence of the electrostatic force used in the calibration. The theoretical calculation of the electrostatic force was verified numerically to better than 1%. The measured Casimir force was shown to increase by 15% at 130 nm separation when the orientation angle between corrugations increased from 0° to 2.4°. The measurements were found to be in agreement with theory based on the derivative expansion which includes the diffraction like correlation effects and the real material properties of the Au surfaces. The comparison between experiment and theory were made with error bars representing the 67% confidence level. No fitting parameters were used in the comparison. The role of the correlation effects and material properties were explored by different comparisons to the PFA, which ignores the complex interplay of the boundary geometry. The ratio of the measured Caimir force to that from PFA showed a deviation as large as 7.7% for a crossing angle of $0^o$ at the shortest separation even for the shallow corrugations used. The deviations due to the coupled correlation and material properties were also examined by comparing the difference of the measured force and PFA to the difference of the theoretical force from the derivative expansion and PFA. The agreement between the experimental difference force and that of the derivative expansion theory with real material properties are included, demonstrate the interplay of the correlation effects of the geometry with the dielectric properties of the boundary. The role of the material properties was independently assessed by a comparison to the derivative expansion theory using ideal metals. The role of temperature in the measured force was studied and the experimental results were shown to be more consistent with the derivative expansion theory at 300 K. The results give an experimental verification of the derivative expansion approach to calculating the Casimir force. These results indicate that the angle dependent Casimir force for two oriented corrugations is an important system for understanding the non-trivial combined interactions of geometry, material properties and temperature. This demonstration of the normal Casimir force between corrugated surfaces will find applications in adjusting and controlling the functionality of closely spaced moving parts of micromachines in the nanotechnology industry.

**ACKNOWLEDGEMENTS**

The authors acknowledge extensive discussions with M. Kardar and G. Bimonte. This work was supported by NSF Grant No. PHY 0970161 (experiment, U.M, A.B), DOE Grant No. DEF010204ER46131 (analysis, A.B, U.M) and NSF grant Grant No. DMR-1310687 (RZ and JW)**.**

**Fig. 1**: Schematic of the experiment setup.

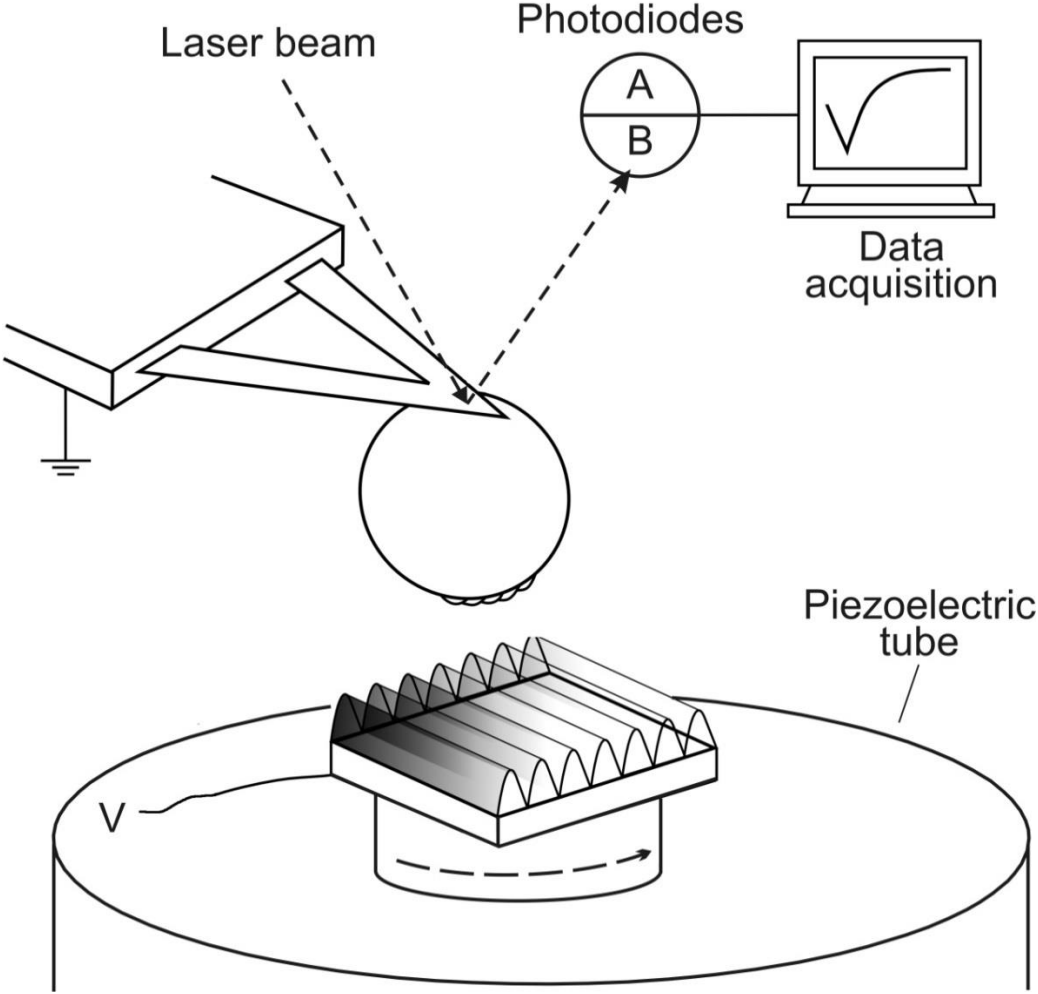

**Fig. 2**: (a) Am AFM scan of the grating surface showing the sinusoidal corrugations. (b) A typical section of the grating surface along y=*const* plane. The solid line is a sine function obtained from the fit.

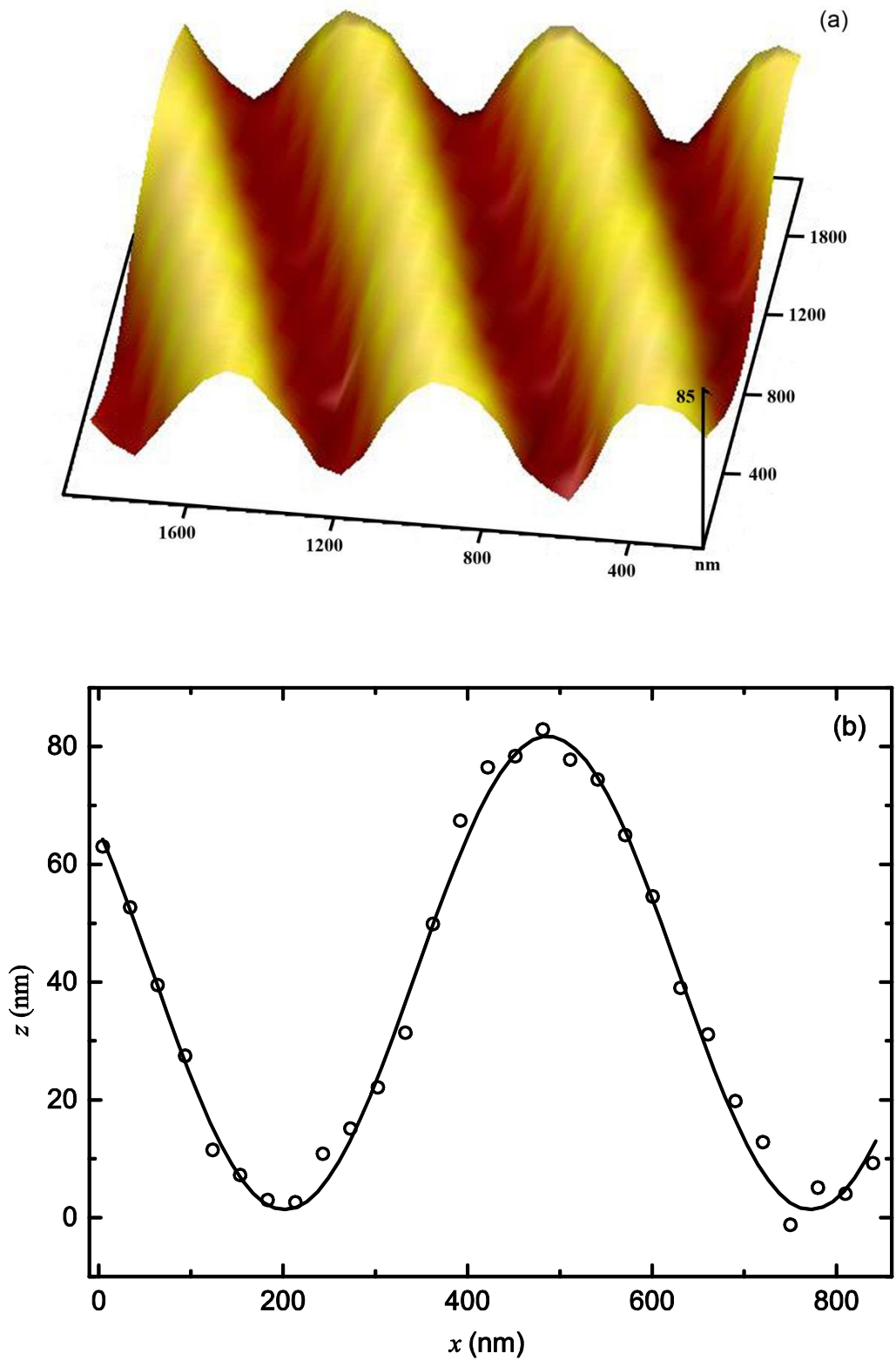

**Fig. 3**: An AFM scan of the surface of the sphere showing the imprinted sinusoidal corrugations. (b) A typical section of the grating surface along a y=*const* plane. The solid line shows a sine function obtained from the fit.

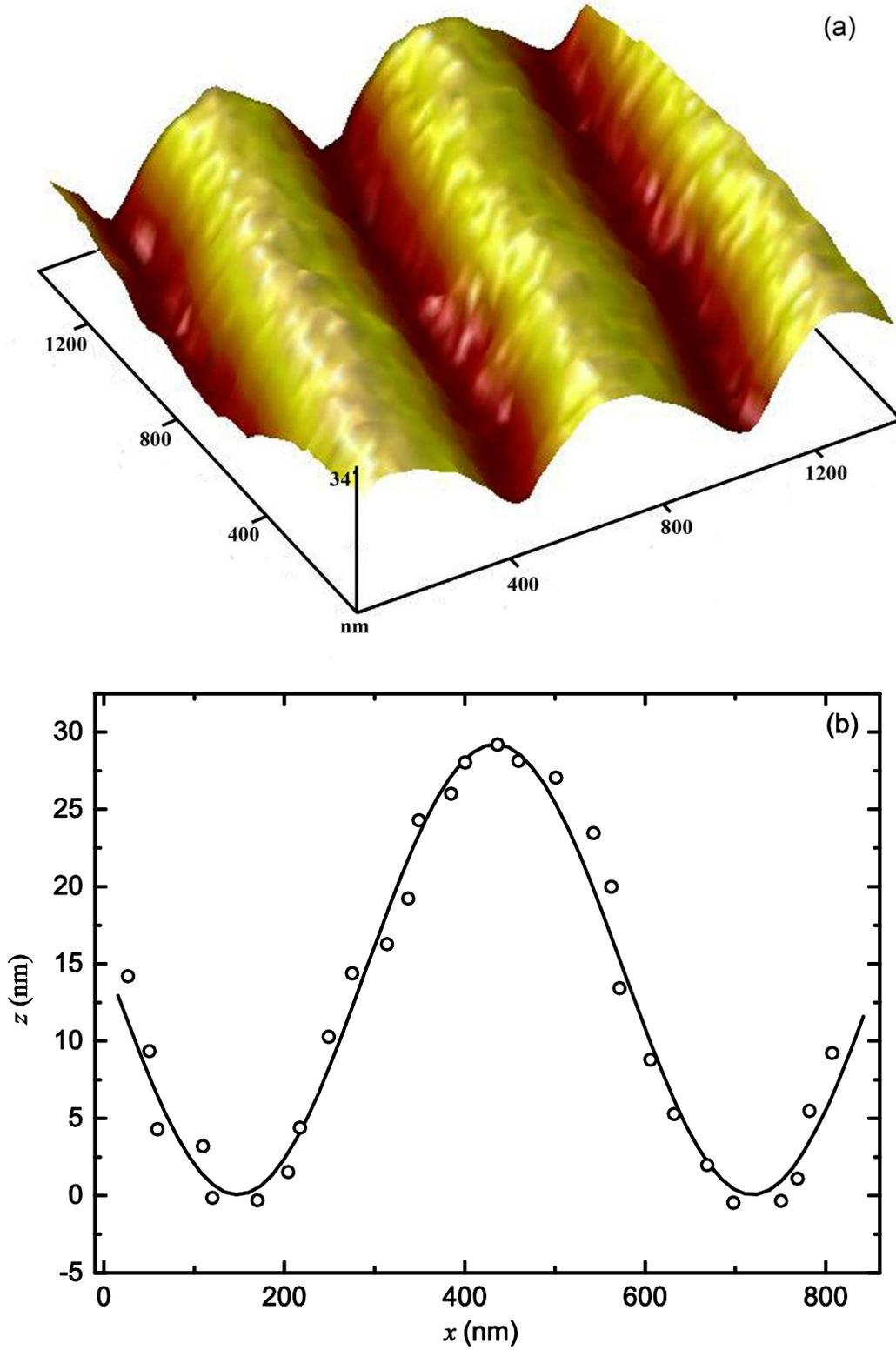

**Fig. 4**: Scanning electron micrograph of the imprint of the corrugations on the sphere.

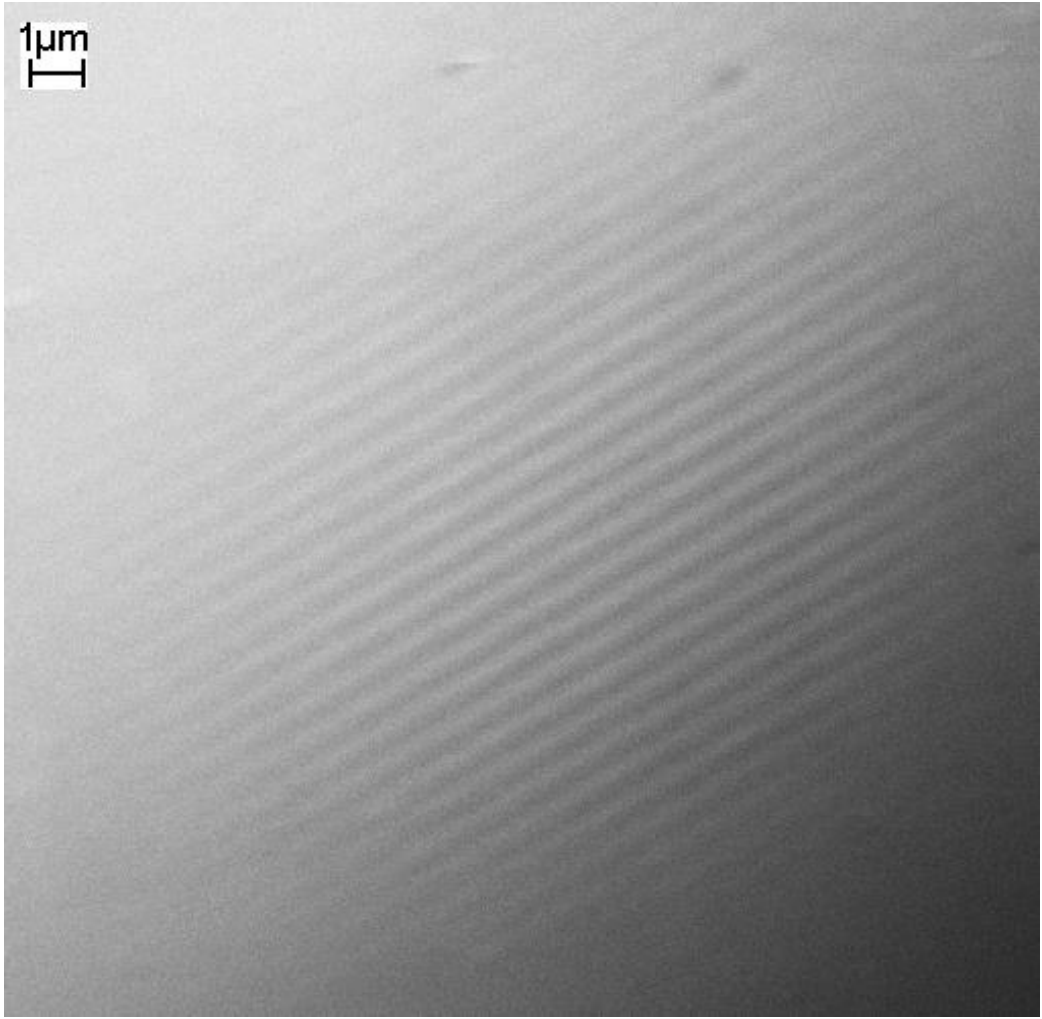

**Fig. 5**. The deflection signal $S_{def}$ as a function of the applied voltage $V$ at a fixed separation of 135 nm between the sphere and the plate. Line is the best fit of the data by parabolic dependence (see Eq. (1)).

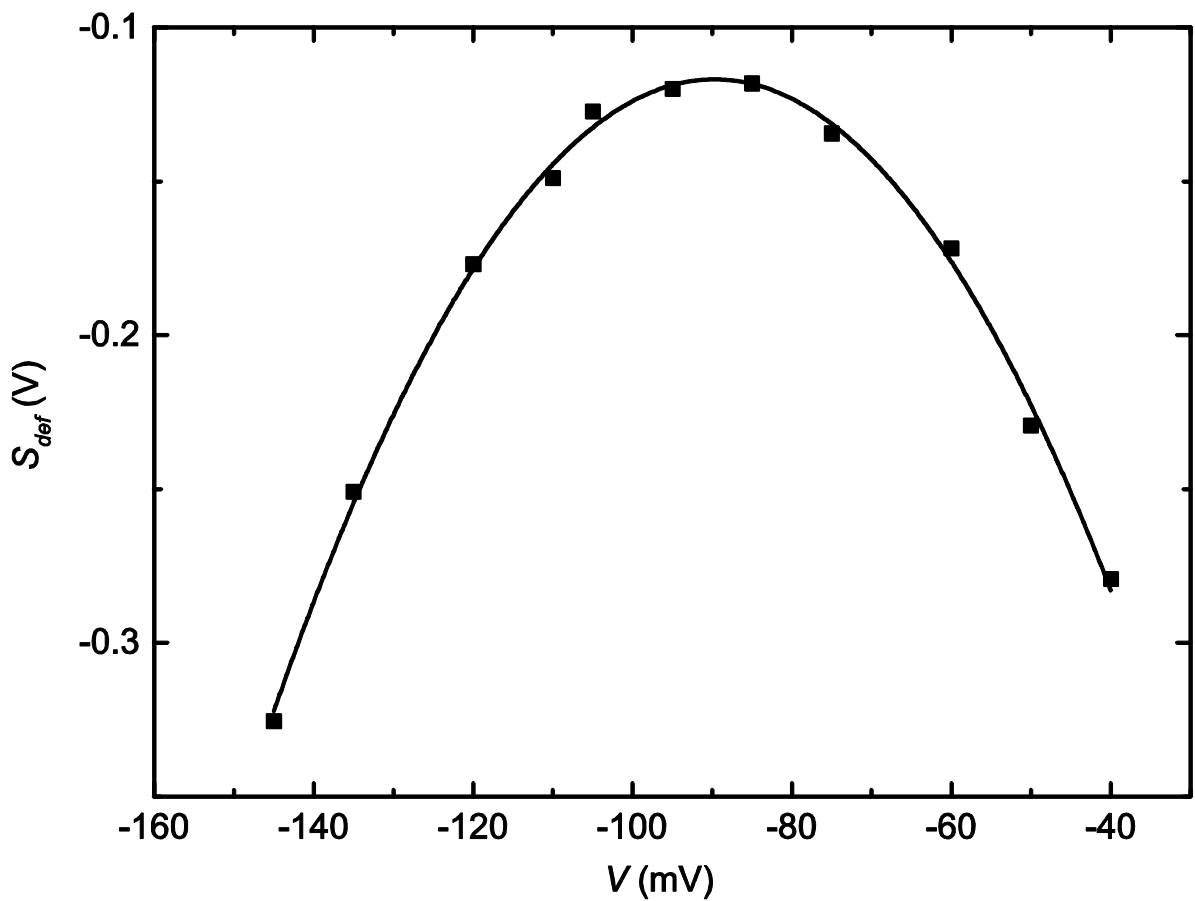

**Fig. 6**. The residual potential difference $V_0$ between the sphere and the plate surfaces as a function of separations for (a) 0, (b) 1.2, (c) 1.8 and (d) 2.4 degree crossing angle between corrugations.

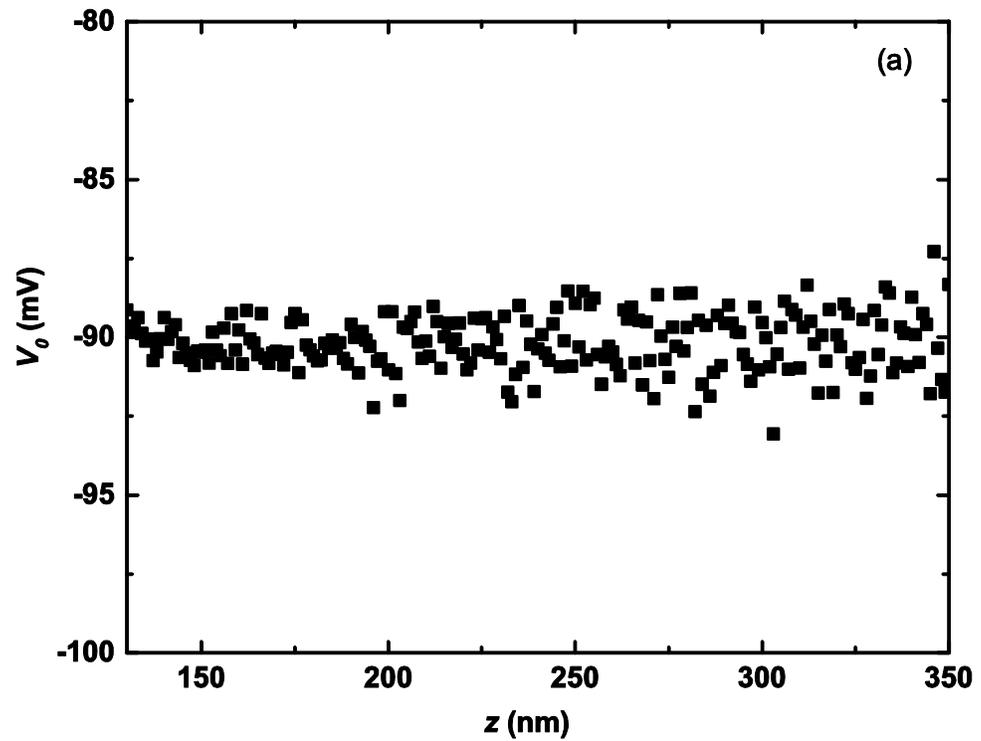

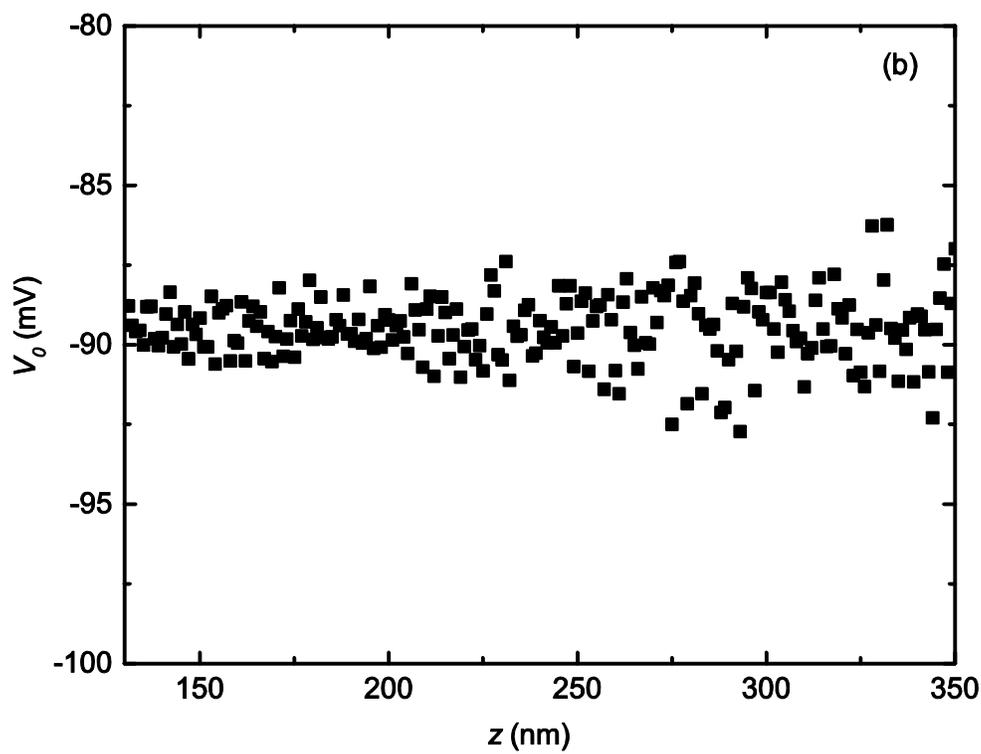

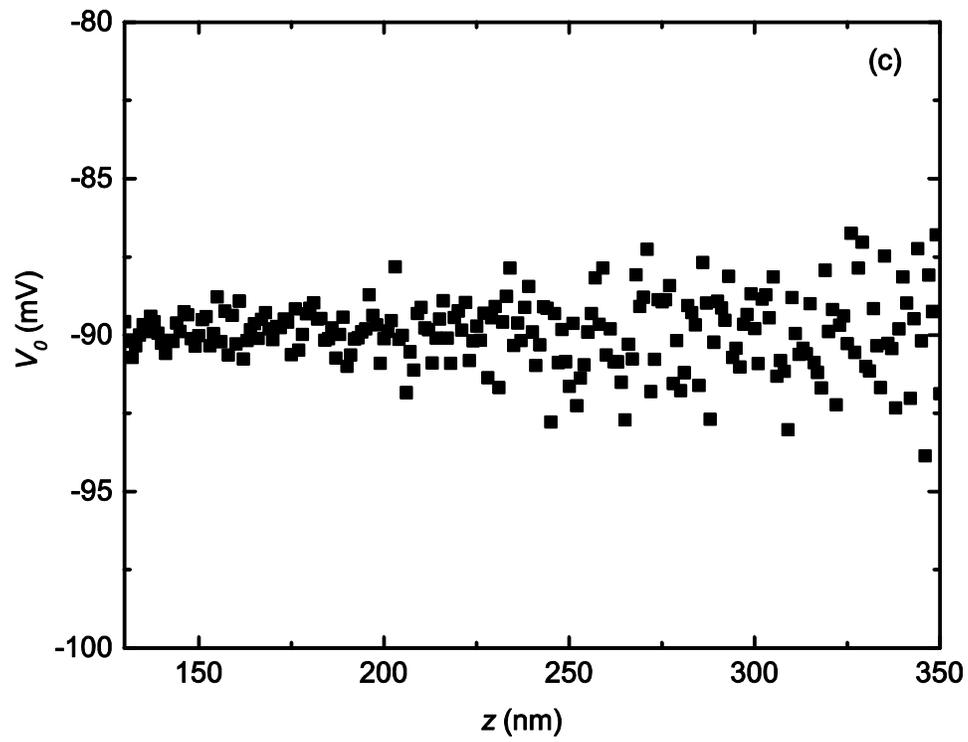
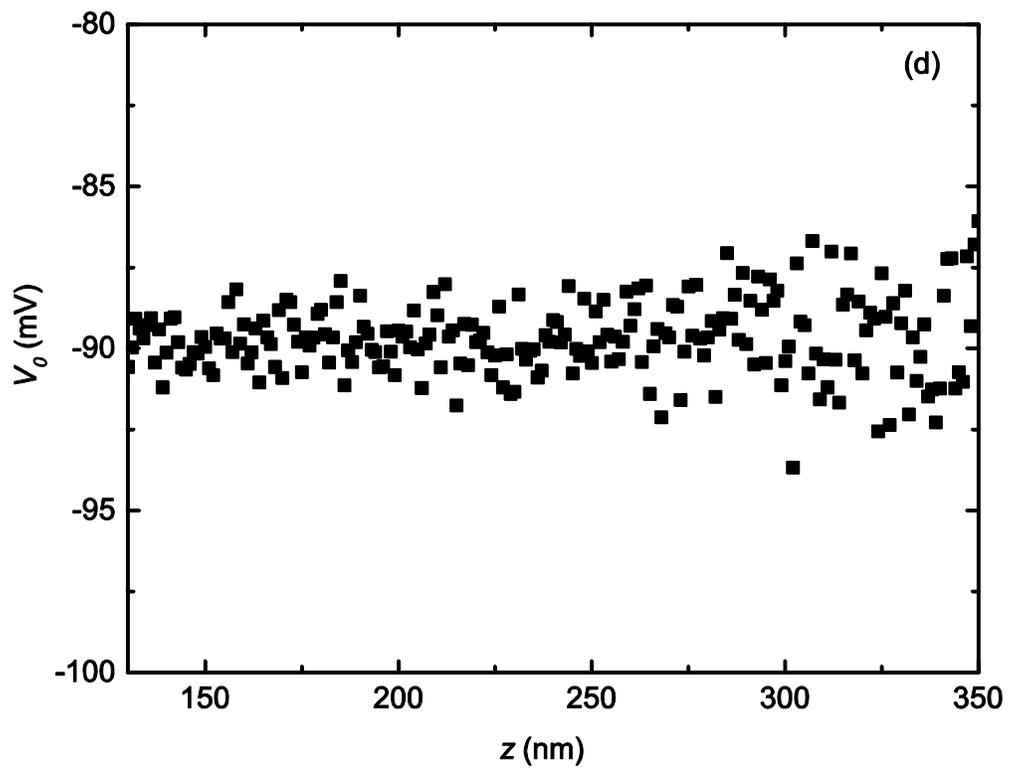

**Fig. 7**. The separation on contact $z_0$ between the sphere and the plate surfaces as a function of the end point $z_{end}$ for (a) 0, (b) 1.2, (c) 1.8 and (d) 2.4 degree crossing angle between corrugations.

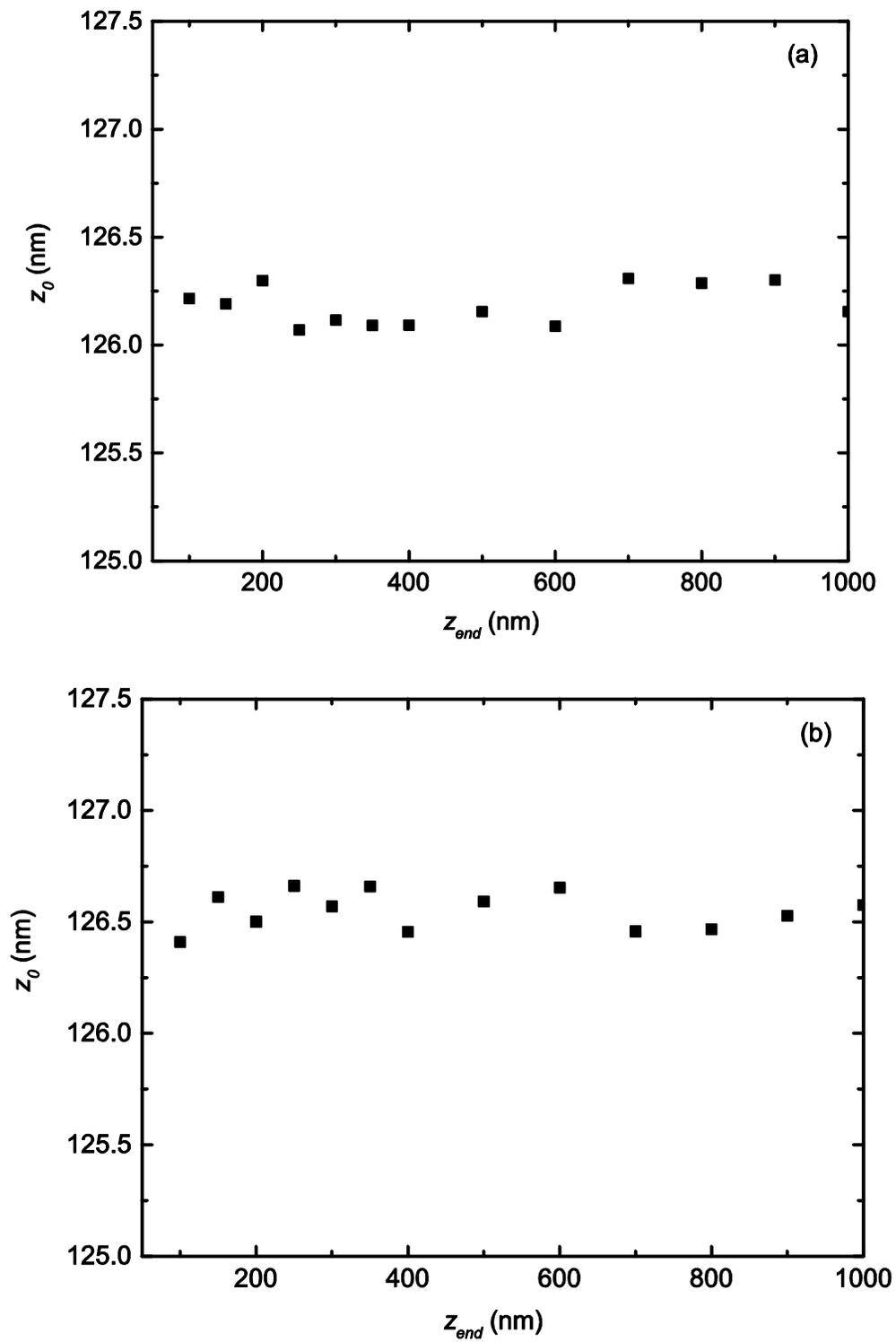

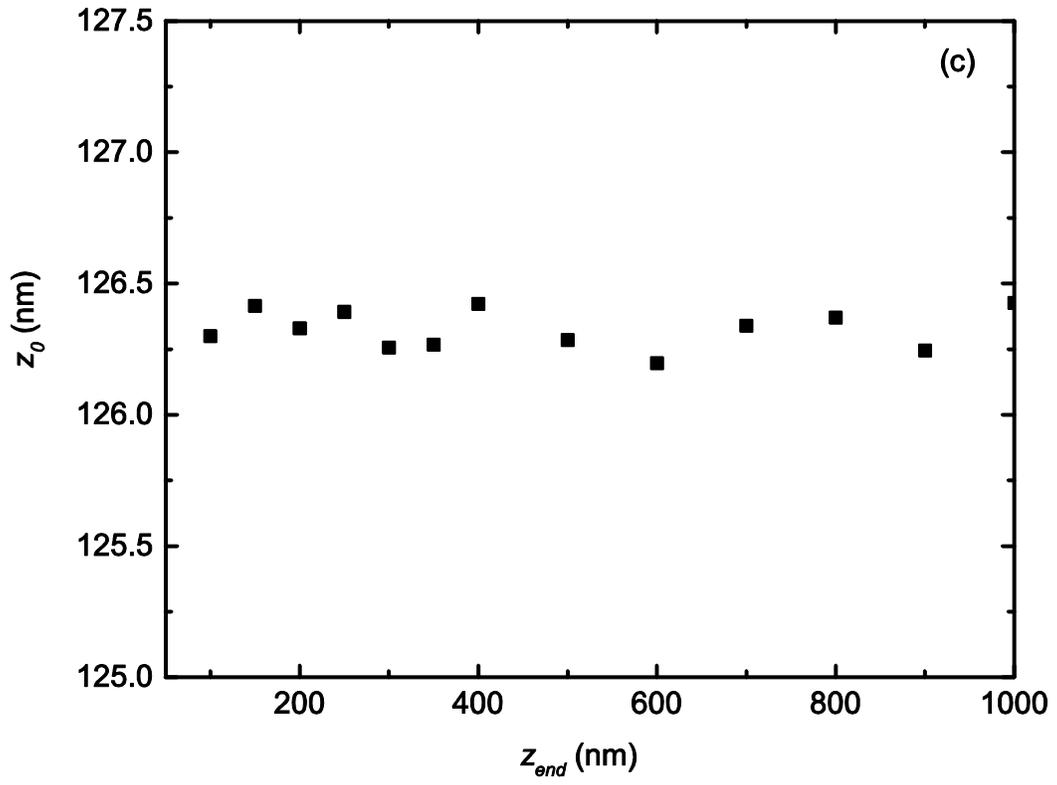
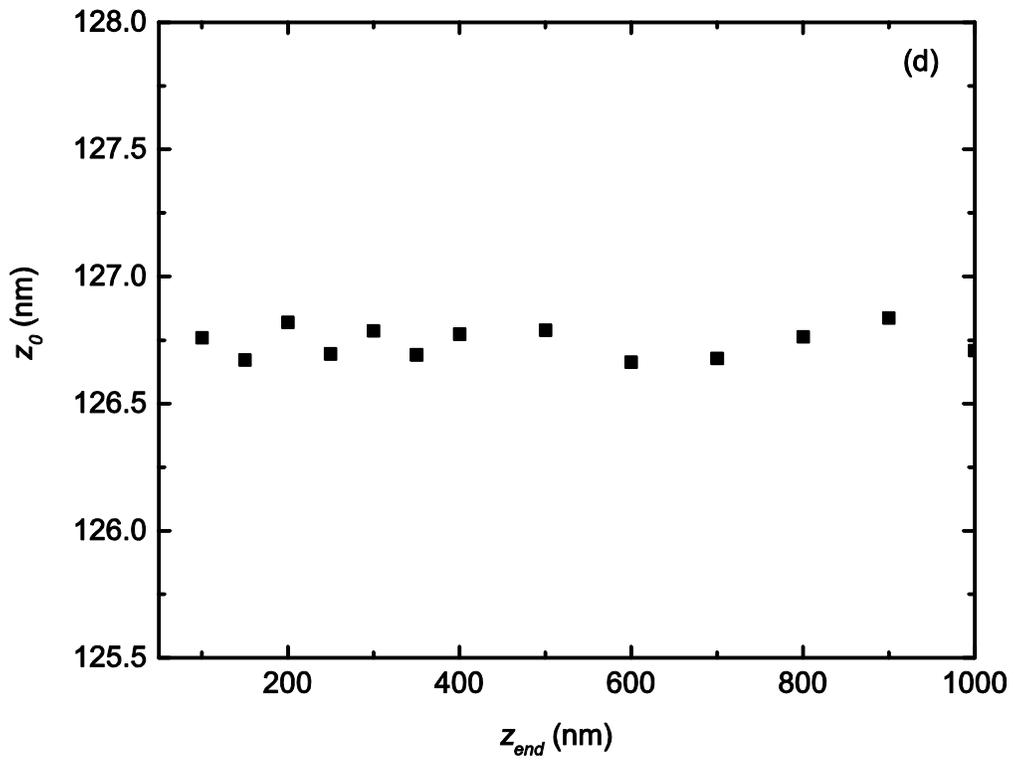

**Fig. 8**. The calibration constant $\sigma'$ as a function of the end point $z_{end}$ for (a) 0, (b) 1.2, (c) 1.8 and (d) 2.4 degree crossing angle between corrugations.

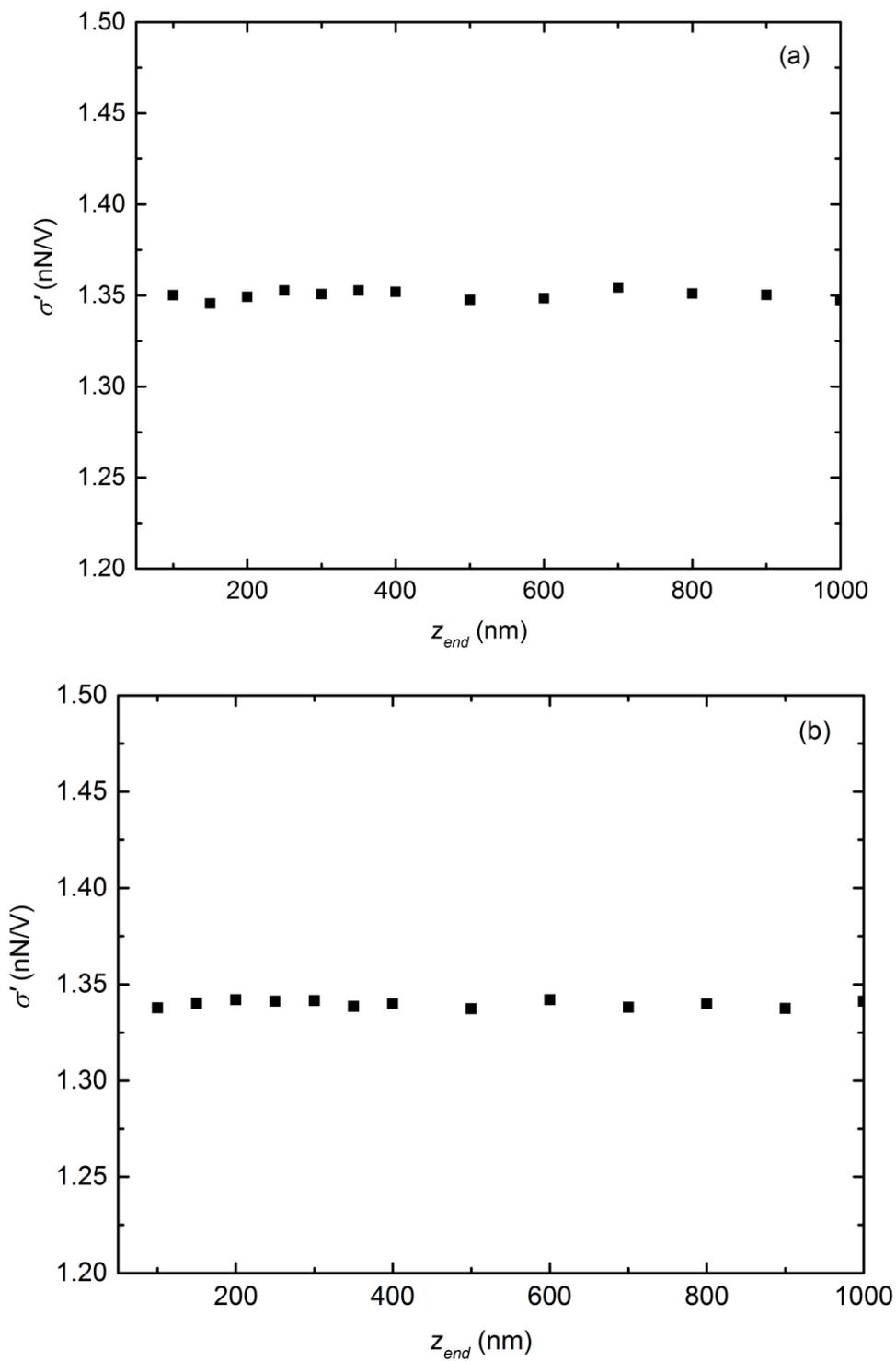

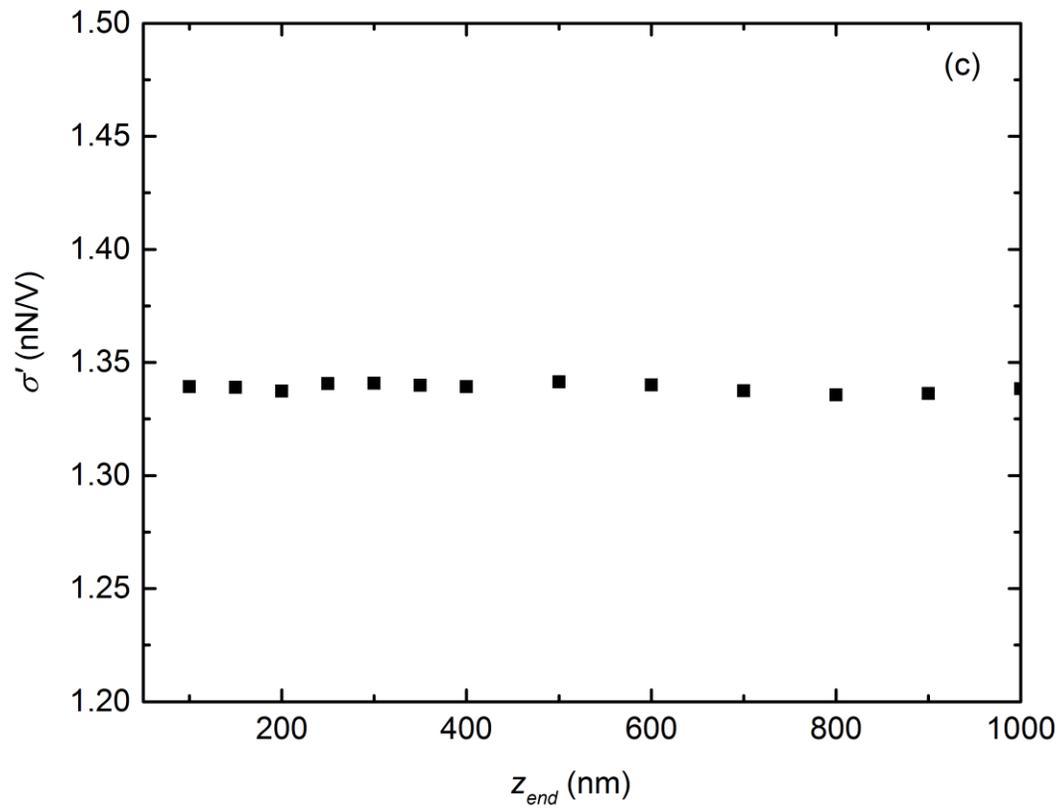

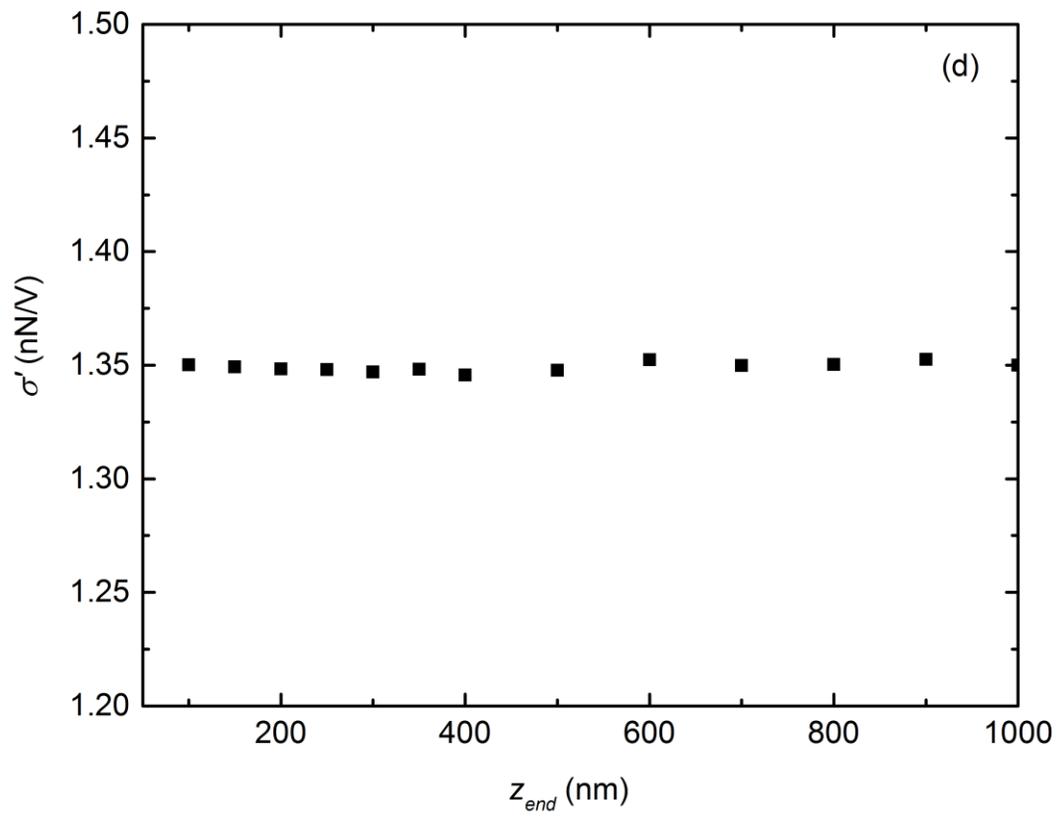

**Fig. 9**: The mean values of measured Casmir forces are shown as crosses for different orientation of the corrugations. The forces from the top to the bottom correspond to orientation angles of $0°$, $1.2°$, $1.8°$ and $2.4°$ respectively. The size of the crosses represents the total error at 67% confidence level. Separations distances from 127 to 230 nm are shown. The solid lines represent the theoretical Casimir forces $F^{Der}$ calculated in section V using derivative expansion. No fitting parameters are used in the comparison of theory and experiment.

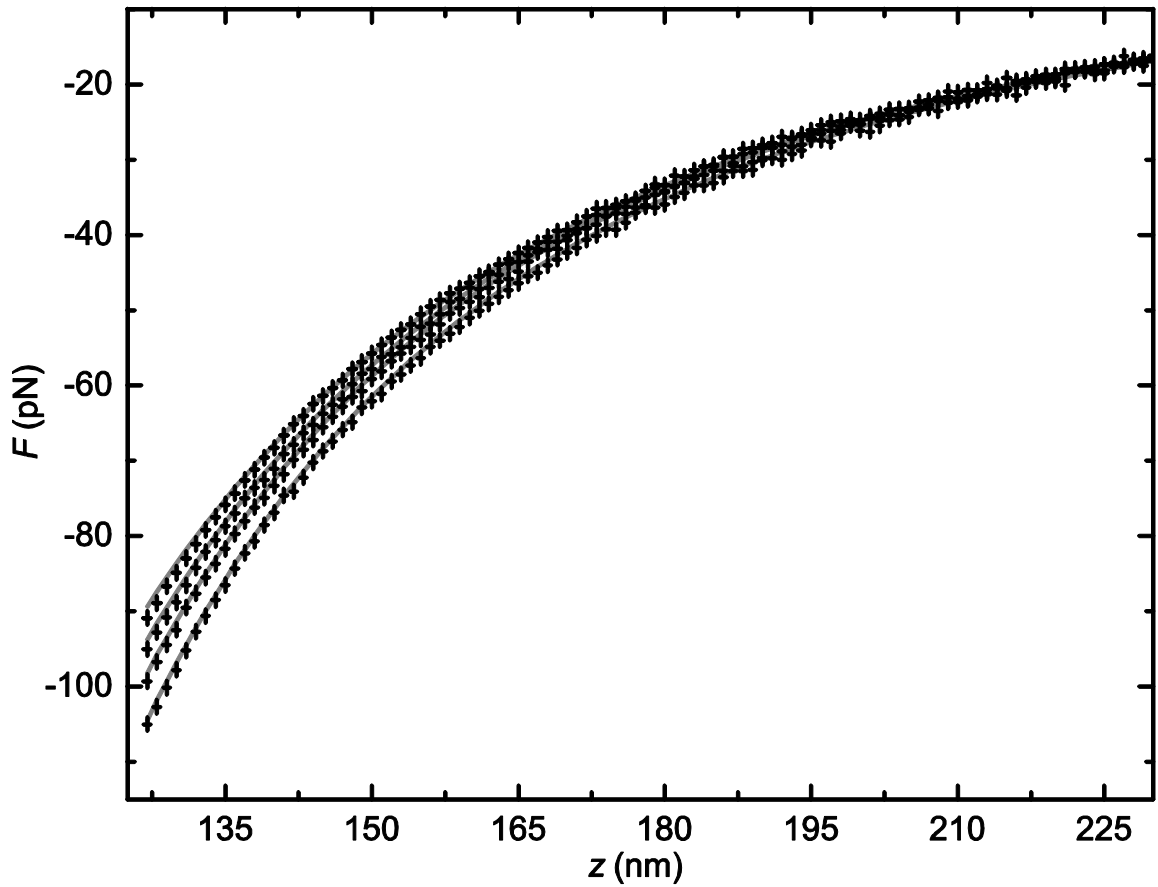

**Fig. 10**: The random (dash line), systematic (dotted line), and total (solid line) errors in the measured Casimir force determined at a 67% confidence level are shown as functions of separation *z* for the 1.2° crossing angel.

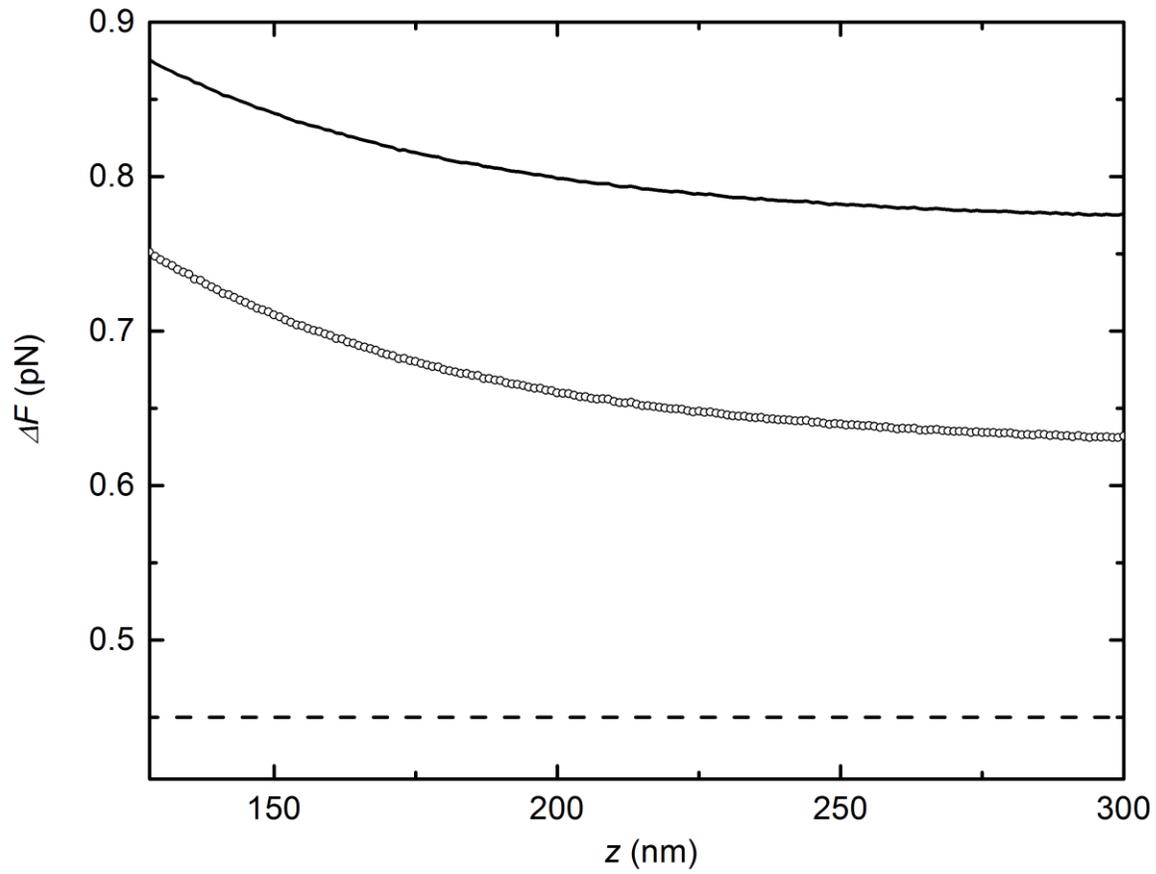

**Fig. 11**: The ratio of measured Casimir forces (presented in the Fig. 9) to the force calculated using PFA at (a) $\theta=0°$, (b) $\theta=1.2°$, (c) $\theta=1.8°$ and (d) $\theta=2.4°$.

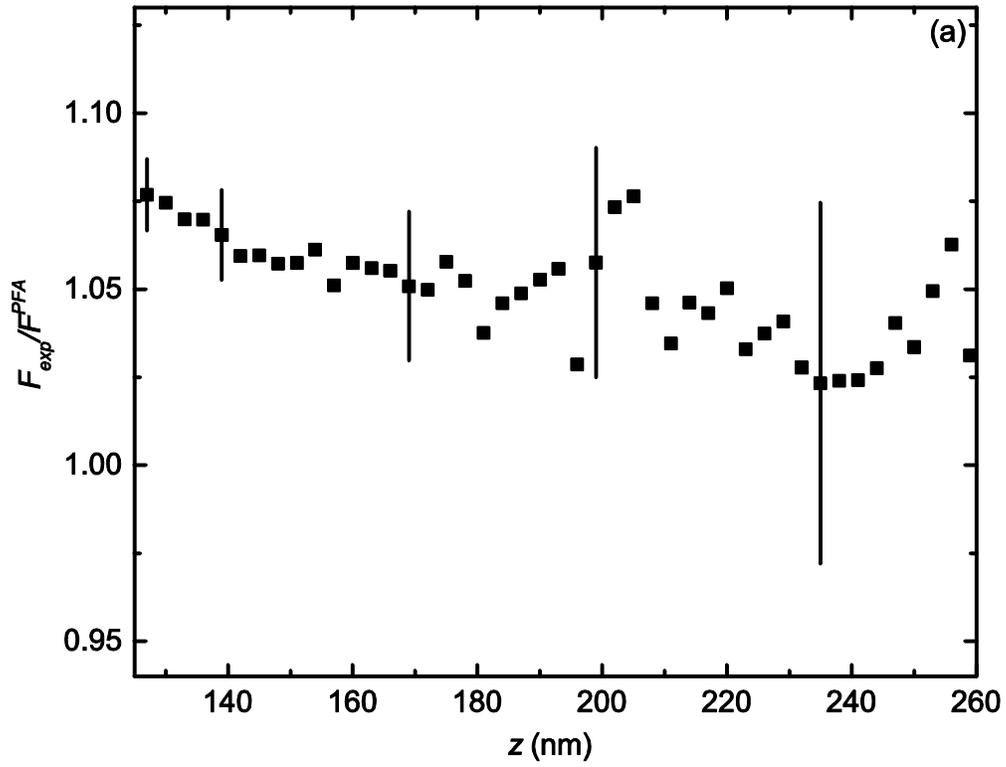

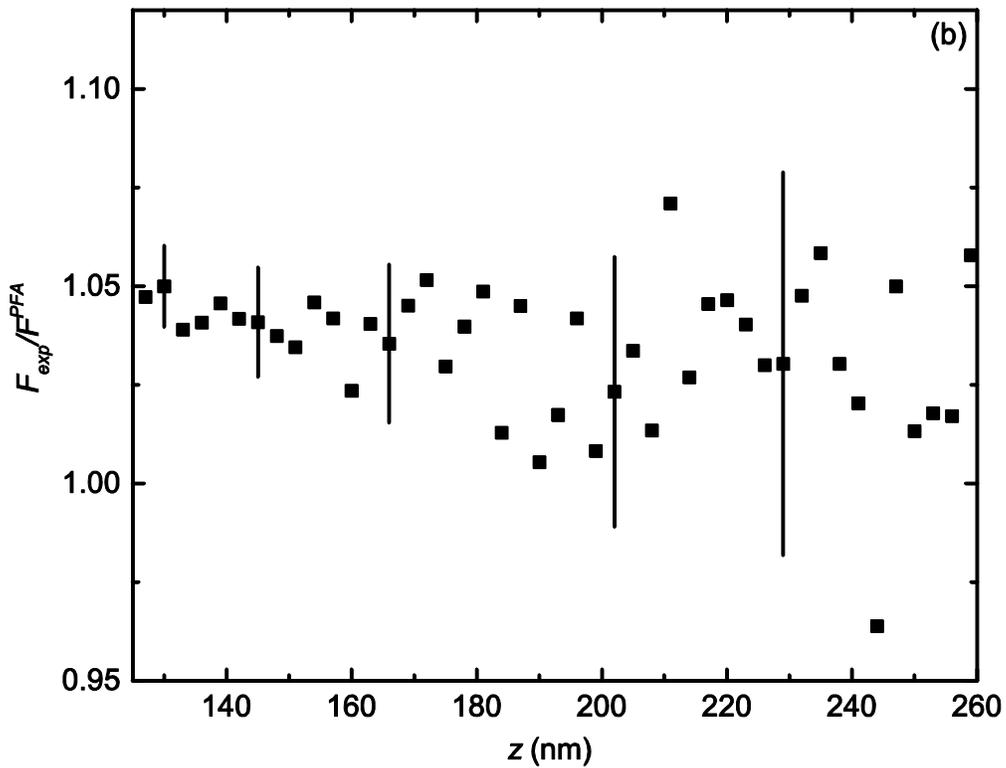

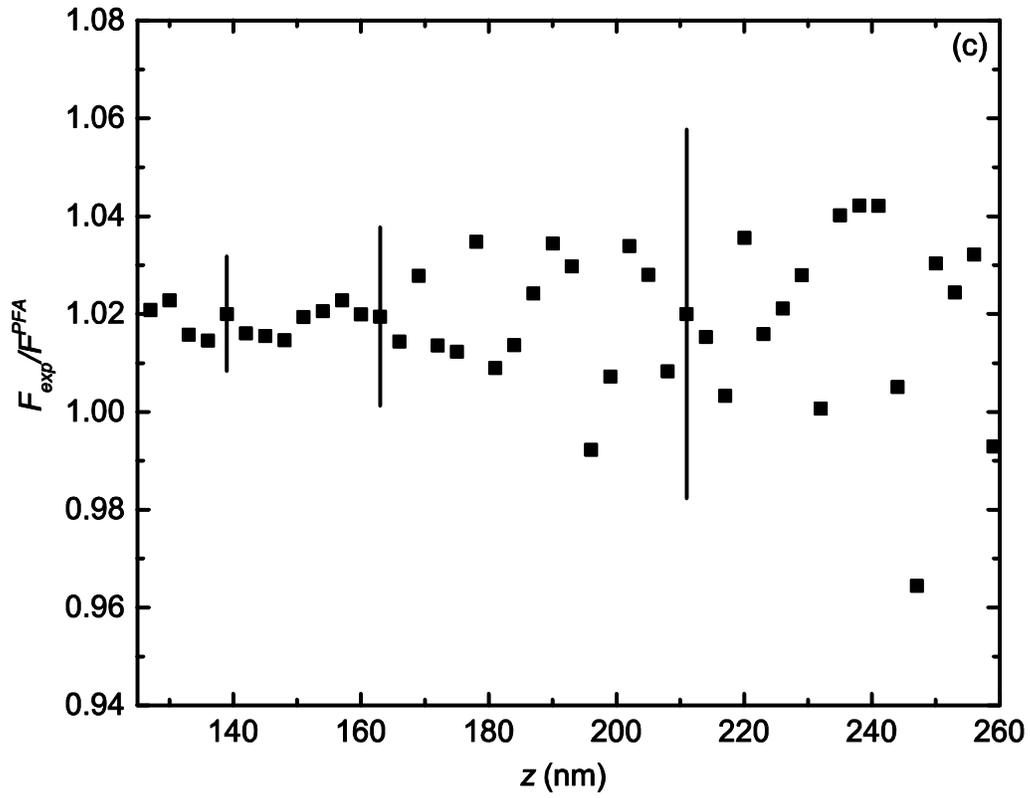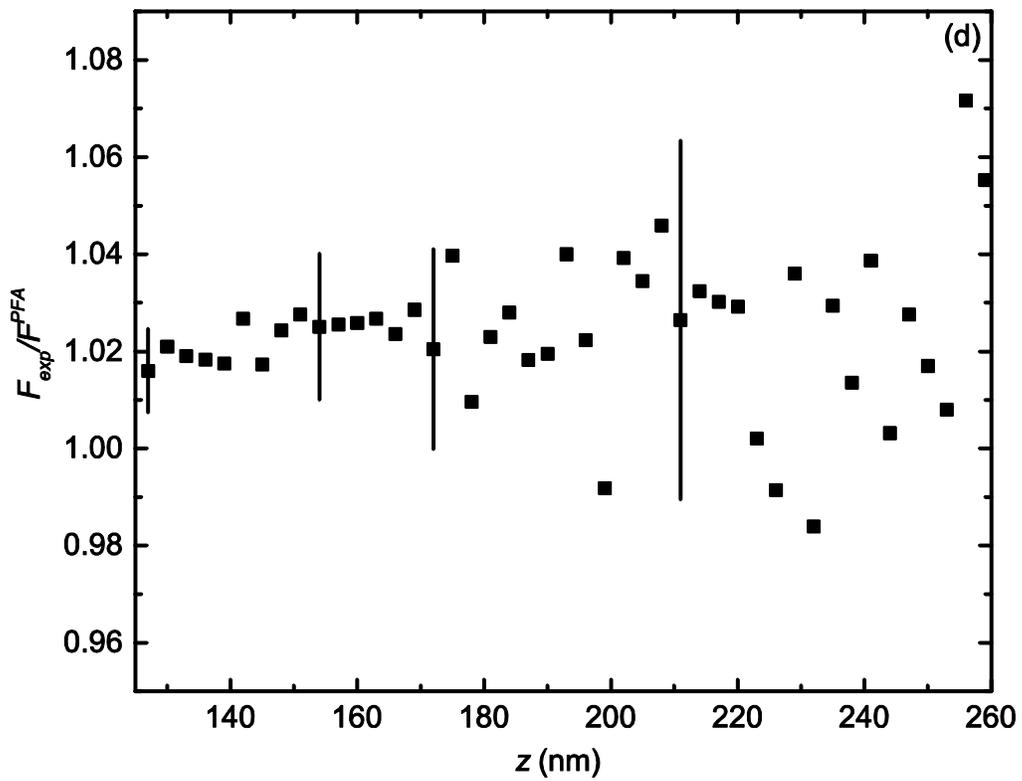

**Fig. 12**: The difference Casimir force $\Delta F_{Cas}=F_{exp}-F^{PFA}$ represented as crosses corresponding to error bars at 67% confidence lever for corrugation orientation angles of (a) $0°$, (b) $1.2°$, (c) $1.8°$ and (d) $2.4°$. The solid line is the corresponding difference between the two theories $F^{Der}-F^{PFA}$ calculated in section V, which is a measure of the correlation effects. Significant deviation from the PFA is observed and the good agreement with the theory based on derivative expansion is found with no fitting parameter. The dashed line is the theoretical difference for ideal metal corrugated surfaces at 300 K. The data is presented every 3 nm for clarity. Inset in (a) and (b) shows the ratio of the data to the force from the derivative expansion at 300 (black squares) and 0 K (gray circles); the same ratios for the angels $1.8°$ and $2.4°$ were indistinguishable.

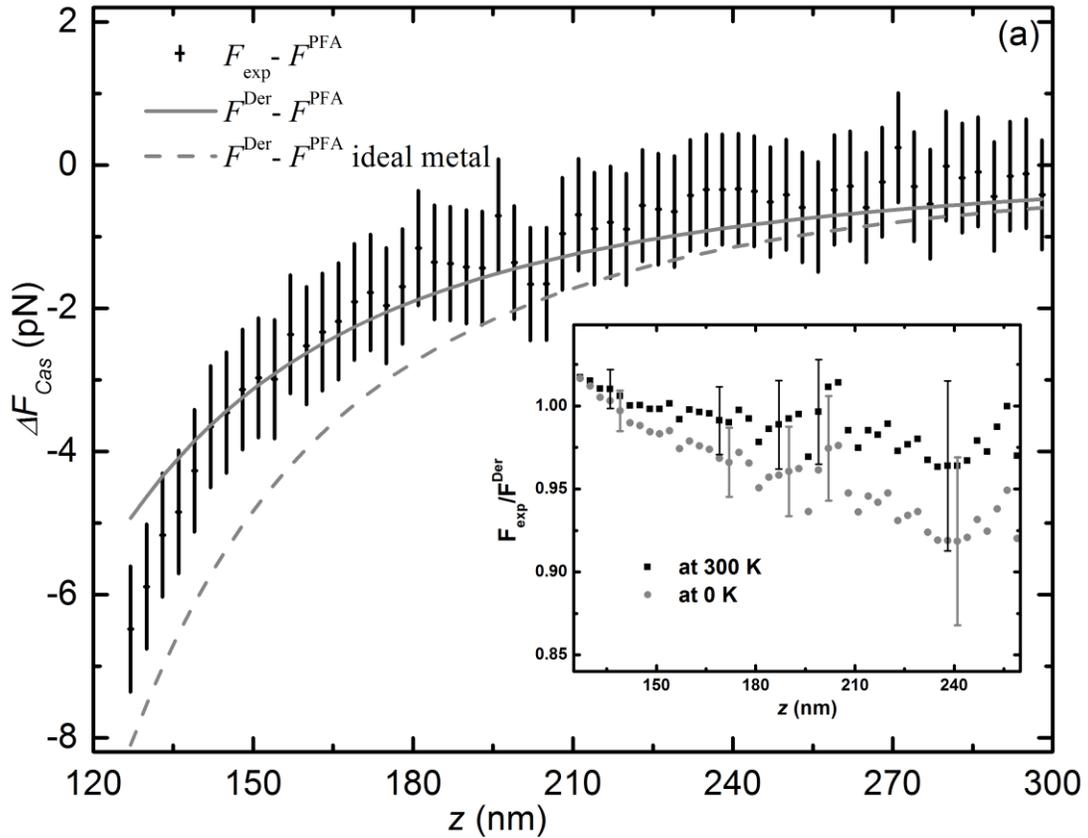

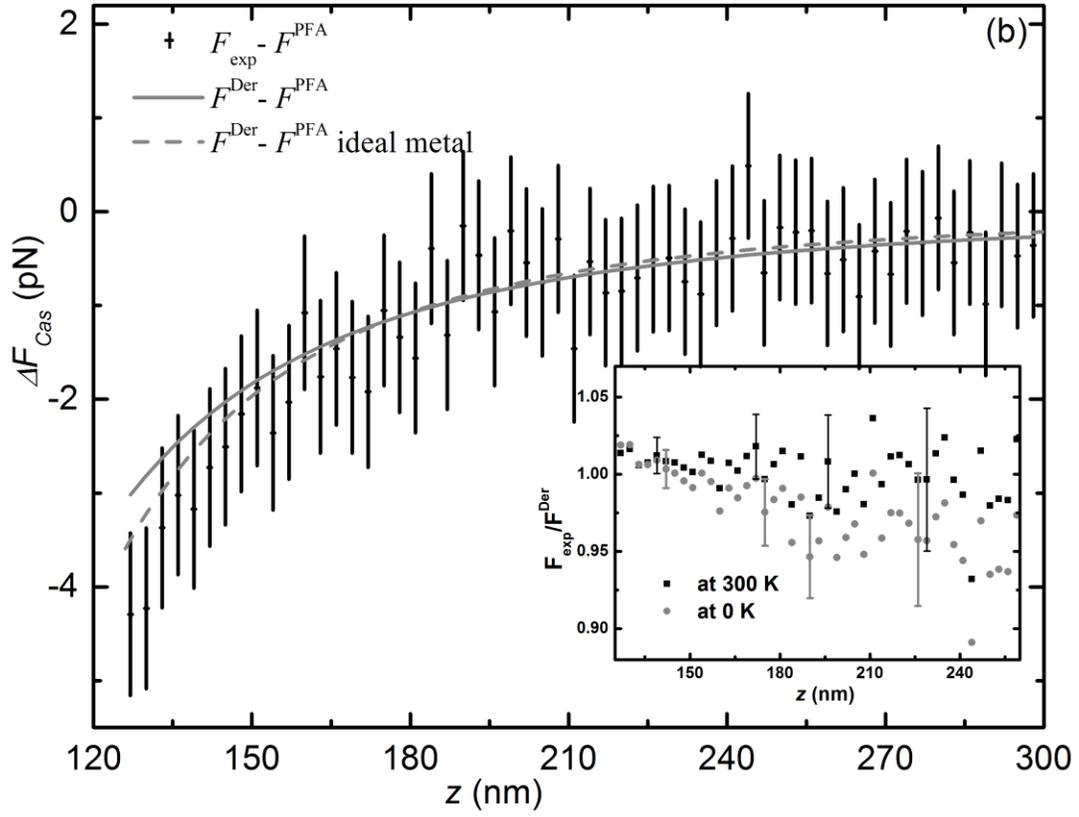

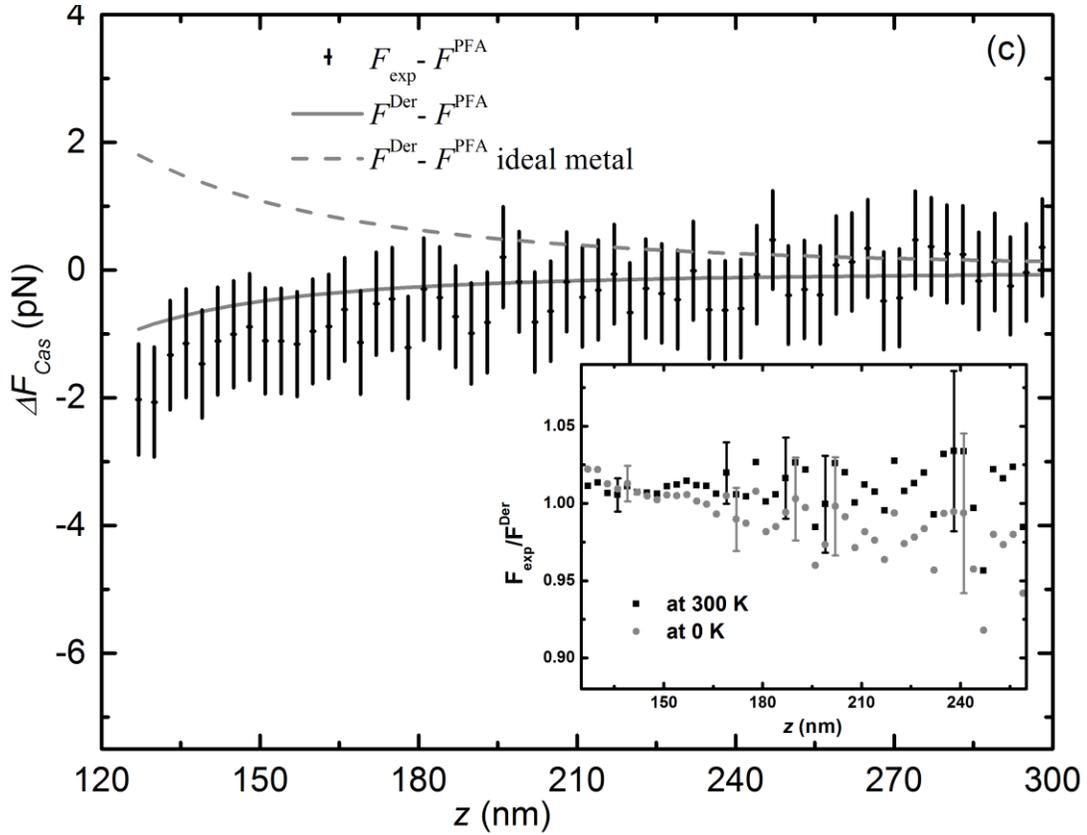

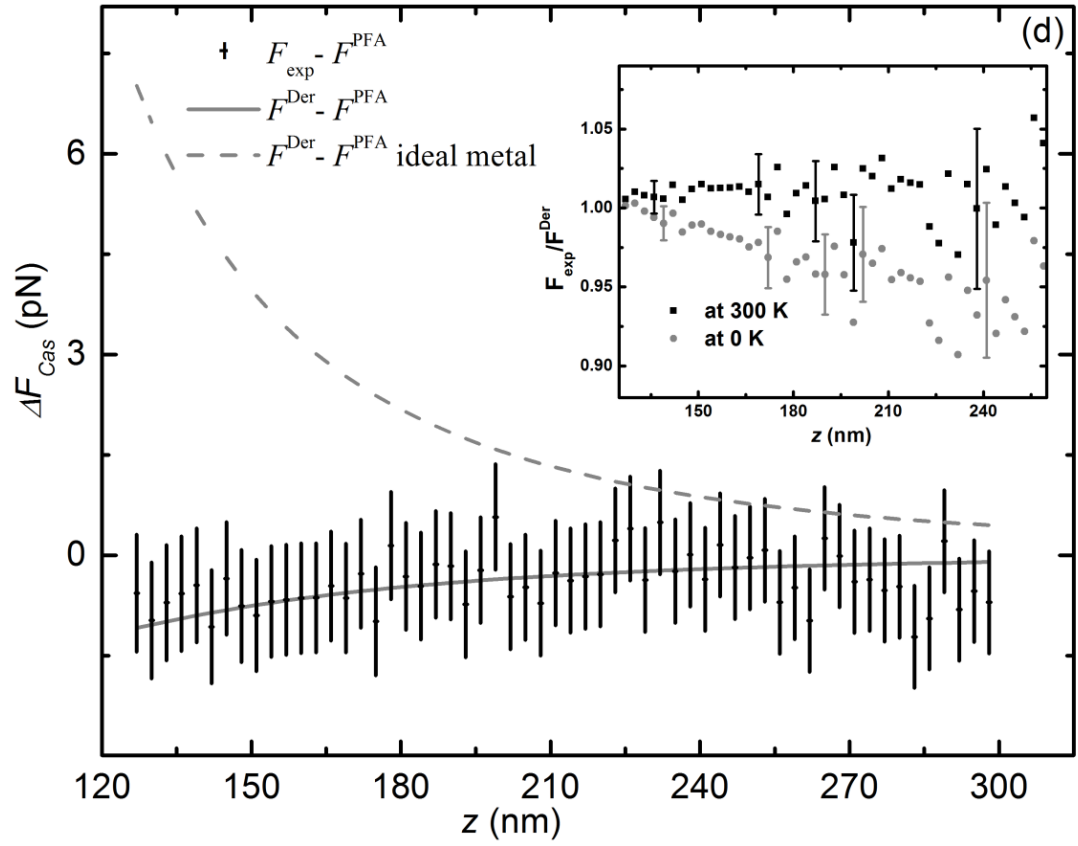

**Table I**. The mean value of residual electrostatic potential $V_0$, the closest separation distance $z_0$ and the cantilever calibration constant $\sigma'$ for each measured crossing angel $\theta$.

| $\theta°$ | $V_0$ (mV) | $z_0$ (nm) | $\sigma'$ (pN/mV) |
|---|---|---|---|
| 0 | -90.2±1.3 | 126.2±0.4 | 1.35±0.02 |
| 1.2 | -89.5±1.1 | 126.5±0.4 | 1.34±0.02 |
| 1.8 | -89.9±1.3 | 126.3±0.4 | 1.34±0.02 |
| 2.4 | -89.7±1.2 | 126.7±0.4 | 1.35±0.02 |